# Causes and Electoral Consequences of Political Assassinations: The Role of Organized Crime in Mexico

Roxana Gutiérrez-Romero♣ and Nayely Iturbe

## Abstract

Mexico has experienced a notable surge in assassinations of political candidates and mayors. This article argues that these killings are largely driven by organized crime, aiming to influence candidate selection, control local governments for rent-seeking, and retaliate against government crackdowns. Using a new dataset of political assassinations in Mexico from 2000 to 2021 and instrumental variables, we address endogeneity concerns in the location and timing of government crackdowns. Our instruments include historical Chinese immigration patterns linked to opium cultivation in Mexico, local corn prices, and U.S. illicit drug prices. The findings reveal that candidates in municipalities near oil pipelines face an increased risk of assassination due to drug trafficking organizations expanding into oil theft, particularly during elections and fuel price hikes. Government arrests or killings of organized crime members trigger retaliatory violence, further endangering incumbent mayors. This political violence has a negligible impact on voter turnout, as it targets politicians rather than voters. However, voter turnout increases in areas where authorities disrupt drug smuggling, raising the chances of the local party being re-elected. These results offer new insights into how criminal groups attempt to capture local governments and the implications for democracy under criminal governance.

♣ Corresponding author: Professor Roxana Gutiérrez-Romero, Queen Mary University of London. r.gutierrez@qmul.ac.uk.

# 1. Introduction

Criminal organizations infiltrate governments to further their agendas, significantly affecting many countries in Latin America, Africa, and Asia, such as Colombia, Brazil, Somalia, Nigeria, Afghanistan, and the Philippines. The influence of these criminal groups often leads to increased political violence and poses substantial challenges to establishing peace and legitimate political authority (ACLED, 2024; Albarracín, 2018). Mexico illustrates the vulnerabilities exploited by organized crime. Drug traffickers, through bribery and violence, have established a form of criminal governance known as narcocracy (Andreas & Youngers, 1989). In response to escalating drug-trafficking violence, President Felipe Calderón launched a war on drugs in 2006, deploying the army to areas with high criminal activity and arresting several drug lords. These crackdowns fragmented criminal organizations, leading to territorial battles and over 300,000 homicides to date. This pervasive violence has also affected politics, with unclear causes and impacts on voting behavior.

In this article, we examine the causes behind the alarming increase in assassinations of political candidates and mayors in Mexico, which has led to nearly 500 politicians being murdered during 2000–21. We also explore the implications of these assassinations for voting behavior. Earlier theoretical studies have investigated how criminal groups use violence as a strategic political tool to capture governments (Alesina et al., 2019; Dal Bó et al., 2006). Empirical studies focused on political assassinations during the first wave of Mexico's drug war, 2006–12, have found links to inter-cartel violence, though the specific drivers and electoral implications remain unclear (Blume, 2017; Hernández Huerta, 2020; Ponce et al., 2022; Rios, 2012; Trejo & Ley, 2021). We argue that criminal organizations in Mexico assassinate candidates and politicians for two primary reasons. First, to gain control over local governments, particularly in strategically important extractive areas, such as those used for oil theft. Second, in retaliation for government actions that threaten to dismantle or significantly disrupt their activities. We also hypothesize that because criminal political violence targets politicians rather than voters, it has a negligible effect on voter turnout.

We test our hypotheses using the 'Political Assassinations, Intimidation and Actors in Mexico' (PAIAMEX) dataset compiled for this article. This database includes 500 records of political assassinations during 2000–21, covering 69 candidates and people contending to become candidates (pre-candidates), 99 incumbent mayors, and 146 former mayors. We examine two types of government crackdowns on criminal organizations: arrests or killings of members of organized crime, and the destruction of illicit drugs. To address the potential endogeneity in these crackdowns and identify their causal effects, we use a fixed effects Poisson estimator with instrumental variables. Our instruments include the number of Chinese migrants arriving in Mexico in the 1930s, some of whom introduced opium cultivation. This historical migration pattern has been shown to influence the locations where drug trafficking groups operate today (Murphy & Rossi, 2020). Additionally, we use other instruments such as the retail prices of heroin and cocaine in the United States, and local corn prices in Mexico, which influence farmers' decisions about drug cultivation (Dube et al., 2016). These instruments satisfy the exclusion criteria, as they do not directly impact political assassinations but explain why certain areas experience higher drug trafficking activity and more intense government crackdowns.

Our findings reveal that political assassinations in Mexico are driven by two main factors: rent-seeking and retaliatory violence. Criminal groups target political candidates for rent-seeking to control smuggling and oil theft routes, especially during elections and periods of exogenous fuel price hikes that make oil theft more lucrative. This control is essential for drug trafficking organizations, which have evolved in Mexico from mere smuggling operations to a model based on territorial extraction. Oil theft exemplifies this shift, as drug traffickers extract gasoline and diesel from the oil pipeline network to sell on illegal markets, generating significant daily profits comparable to those from drug smuggling (Ferri, 2019). For mayors, the risk of assassination increases in areas closer to oil pipelines if they remain involved in politics or become entrepreneurs after leaving office. This heightened risk is attributed to their potential connections with criminal organizations, as indicated by independent intelligence reports (Mejía 2021).

Our analysis suggests that, in this context, political violence is driven more by criminal motives than by political competition. Specifically, we find that retaliatory violence—triggered by government actions such as arresting or killing organized crime members—significantly increases the likelihood of incumbent mayors being assassinated. Notably, the political affiliations of these mayors, whether aligned with the governor's or the president's party, do not influence the likelihood of their political assassination.

In line with our expectations, political assassinations have a minimal impact on voter turnout in local elections and do not affect the re-election of the incumbent party. In contrast, voter turnout increases in municipalities where authorities disrupt drug smuggling, boosting the chances of the local party being re-elected. We present compelling evidence that this rise in turnout is likely driven by genuine voter support.

## 2. Political assassinations and their electoral impact

'El Narco', the collective noun term used to refer to drug traffickers, dominates societies where the corrupt state ensures their existence, protection, and survival by allowing these groups to establish a type of criminal governance, referred to as a narcocracy. Narcocracies seek to establish criminal supremacy over territory, influence enforcement policies, and exploit as many natural resources as possible, rather than overthrow or replace the state. Initially, the primary source of profit for these criminal organizations is the cultivation and transportation of illicit narcotics (Andreas & Youngers, 1989). As these groups gain more power, they may shift from simply smuggling narcotics to a territorial extraction mode, engaging in extortion, human trafficking, money laundering, kidnappings, illegal logging, mining, and oil theft (Jones & Sullivan, 2019). This is why narcocracies are typically characterized by drug trafficking organizations that possess semi-autonomous cells and control a network of territories outside their home base. Often, no single drug trafficking group operates exclusively in a single territory, increasing the chances of conflict (Durán-Martínez, 2008). However, as witnessed in Mexico, in the few places where just one drug trafficking organization operates, violence may be quite high due to internal fights within the organization and accompanying crimes (Gutiérrez-Romero, 2016).

These criminal groups gain power through bribery, including funding the campaigns of political candidates who can support their interests (Ponce, 2019). However, the state is not a monolithic entity. Pressures from both local and international actors often push for the eradication of drug trafficking organizations. Several theoretical studies have formalized that criminal organizations respond by using both bribes and strategic political violence to influence policy (Alesina et al., 2019; Dal Bó et al., 2006). These groups target their political violence around elections to influence the candidate pool, either by forcibly forming alliances with contending candidates or dissuading rivals. The ultimate objective of such pre-electoral violence is to establish ties with the elected government to influence policy (Albarracín, 2018; Alesina et al., 2019; Daniele & Dipoppa, 2017; Durán-Martínez, 2008; Ponce et al., 2022). Additionally, criminal groups can use post-electoral violence as a form of coercion against politicians who fail to honor agreements or who jeopardize the organization's activities (Acemoglu et al., 2013; Dal Bó et al., 2006). Notable examples include the terrorist attacks in Colombia by Pablo Escobar to prevent extradition of narcos to the United States, and the Sicilian Mafia's targeting of legislators and politicians in Italy (Acemoglu et al., 2013; Alesina et al., 2019; Daniele & Dipoppa, 2017).

To understand the complex relationship between organized crime and political assassinations in Latin America, recent literature has tested three hypotheses: repression, competition, and rent-seeking. The repression hypothesis posits that criminal groups use violence to influence local governments, control smuggling routes, and defend themselves. For instance, Trejo and Ley (2021) find a positive correlation between higher homicides in Mexican municipalities related to state-cartel conflicts and inter-cartel battles and, an increased risk of political assassinations during 2007–12. However, this correlation might mask other factors contributing to the surge in political violence, such as weakened rule of law in certain vulnerable areas.

Similarly, the competition hypothesis asserts that organized crime kills politicians to gain control over territory and influence governments. Supporting this, Rios (2012) finds a 0.02 increase in the likelihood of political assassinations in Mexican municipalities for every 10 homicides (per 100,000 inhabitants) linked to inter-cartel competition. However, Trejo and Ley (2021) argue that this correlation does not clarify the motives or predict occurrences. Other studies link the rise in political assassinations to the number of criminal organizations in Mexican municipalities, based on estimates from media records during 1990–2010 (Hernández Huerta, 2020; Ponce, 2019). Detecting the number of criminal groups operating in municipalities using media records alone is an impossible task today. These groups have increased to well over 150 compared to the initial dozen groups during the earlier period (Reina, 2011). Additionally, this approach does not reveal motives or predict occurrences of political violence.

The rent-seeking hypothesis suggests that criminal organizations finance their operations by extorting rents from local governments, as evidenced in Colombia during the 1990s (Chacon 2018). Similarly, Trejo and Ley (2021) show that Mexican municipalities with higher fiscal income have a 4.5% increased chance of political killings. However, fiscal revenue may be a confounding factor, making these areas more prone to political violence. These fiscal resources are often audited and limited.

## 2.1 Our three hypotheses

Our central argument is that organized crime can resort to political violence for two reasons. First, for *rent-seeking* behavior, to influence which political candidate wins local elections and capture incoming local governments in areas of strategic value for extraction. Second, as *retaliation* for state measures that threaten their survival or economic interests. Since it is more *cost-effective* for these organizations to target politicians rather than voters, we hypothesize that political violence is likely to have a negligible impact on voter turnout, as explained below.

We hypothesize that criminal organizations use *rent-seeking* violence against candidates in valued extractive areas where the benefits outweigh the risks involved, such as arrests. We contend that, for criminal organizations, regardless of whether one or several are operating in the area, it is crucial to control municipal mayors in high-value areas due to their oversight of municipal resources and their role in municipal security. Mayors are in charge of the municipal police and must collaborate with law enforcement to combat crime in their territory while complying with the governor's directives and federal goals. Co-opting mayors thus reduces transaction costs for criminal organizations operating in crucial areas, such as resource extraction, smuggling, and money laundering, while helping them evade prosecution by local police (Ponce, 2019). Additionally, criminal groups may gain access to intelligence operatives in the area (Hernández Huerta, 2020).

Crucial to criminal organizations, mayors are paradoxically the primary targets for political violence orchestrated by these groups. Their vulnerability stems from their modest salaries, around $2,800 dollars monthly in Mexico, and insufficient protection, exacerbated by potential discrepancies in local, state, and federal security strategies. National crackdowns on criminal organizations, intended to cause economic damage or dismantle leadership, often prove futile as these groups are resilient and violently retaliate against adversaries (Calderón et al., 2015).  Time can compensate for economic losses, but time spent in prison cannot. This makes the risk of retaliatory violence likely to intensify even further after arrests. The goal of such violence is not to overthrow the state but to exert influence over policymaking and to punish or eliminate individuals perceived as threats.

Despite the capacity of criminal organizations to target high-profile politicians, the considerable danger of governmental response makes mayors the most probable targets for criminal organizations seeking retaliation or pressure. This pattern is evident in Italy, where local officials confront Mafia violence aimed at influencing legislation (Daniele & Dipoppa, 2017). In Mexico, intelligence assessments indicate that former mayors have been targeted and even killed because politicians who were once co-opted by organized crime become their economic partners (Mejía, 2021). The absence of legal mediation mechanisms for illegal activities increases the likelihood of these mayors being killed, even years after completing their term.

Organized crime, with ample resources, can also resort to electoral violence to directly influence voters, but this strategy is risky and can potentially lead to their arrest. Moreover, the effectiveness of intimidating voters is uncertain due to the secrecy of the vote. If alternatives exist, voters may reject candidates associated with criminals. A more cost-effective approach for criminal organizations is to use bribery and violence to directly shape the candidate pool. However, a crucial question persists: How does violence against politicians impact voting behavior?

Internationally, the impact of pre-electoral violence on turnout and vote choice is inconclusive (Gutiérrez-Romero & LeBas, 2020). For instance, in Brazilian favelas, support for police-affiliated candidates has increased as a response to combat criminal groups (Hidalgo & Lessing, 2015). In contrast, Alesina et al. (2019) find that violence in Italy reduces voting for parties opposed by criminal organizations. In Mexico, limited evidence exists on the impact of the recent surge in candidate assassinations on political participation. For instance, Ley (2017) finds that areas with higher homicide rates and candidate assassinations experienced lower turnout during the 2012 elections. Despite the ongoing pre-electoral violence targeting candidates, Mexico has witnessed two peaceful democratic alternations in the presidency since the war on drugs began, reflecting a desire for a change in security strategy. Given the significance of elections in Mexico and the fact that violence targets politicians rather than civilians, we expect that the voter turnout in local elections will not be significantly impacted. This discussion leads to the following hypotheses.

Rent-seeking hypothesis

**H1A**: Areas with a competitive advantage for illegal rent extraction experience more political assassinations of candidates contending in local elections than other areas.

**H1B**: Areas with a competitive advantage for illegal rent extraction experience more political assassinations of former mayors than other areas.

Retaliation hypothesis

**H2**: Areas where government actions pose a risk of dismantling or significantly disrupting the activities of criminal organizations experience more political assassinations of incumbent mayors in retaliation compared to other areas.

Cost-effectiveness hypothesis

**H3A**: Areas with a competitive advantage for illegal rent extraction do not experience more electoral violence targeted at voters than other areas.

**H3B**: Areas where government actions pose a risk of dismantling or significantly disrupting the activities of criminal organizations, do not experience more electoral violence targeted at voters than other areas.

**H3C**: Areas that experience political assassinations see a negligible to small decrease in voter turnout compared to other areas.

## 3. Setting

During Mexico's Partido Revolucionario Institucional (PRI) rule from 1929 to 2000, drug trafficking organizations (DTOs) forged alliances with government and enforcement agencies to conduct illegal operations in designated areas, known as *plazas*. The arrangement was simple: in exchange for bribes, DTOs smuggled drugs into the United States, avoided conflicts over other plazas, and refrained from inciting violence. The government also regulated where criminal organizations could cultivate illicit drugs, while personnel of the army, federal, and municipal police protected drug cargoes, charging a $60 dollar tax per kilo (Hernández, 2014).

In the 1960s, pressured by the United States, the Mexican government received significant subsidies to combat drug trafficking, leading to militarized anti-drug campaigns in the 'golden triangle' of Chihuahua, Sinaloa, and Durango. In a partly real, partly simulated effort, the army routinely destroyed illicit cultivations and apprehended minor traffickers. However, instead of eliminating drug trafficking from the region, these operations merely spread it to Guadalajara. In the 1980s, cocaine emerged as one of the fastest-growing and most profitable industries globally (Andreas & Youngers, 1989). Facing crackdowns on the Caribbean route, Colombian cartels turned to Mexico, smuggling 60% of cocaine into the United States. Corruption and innovative smuggling methods facilitated this shift. CIA agent Barry Seal, initially supporting the Contras in Nicaragua, began transporting cocaine from the Medellín Cartel to Mexico's Guadalajara Cartel, led by Miguel Ángel Félix Gallardo, Ernesto Fonseca Carrillo, and Rafael Caro Quintero. In return, the Medellín Cartel funded the Contras and paid Seal $1.5 million per trip (Hernández, 2014). Despite Seal's assassination in 1986, cocaine, marijuana, and heroin smuggling from Mexico to the United States persisted.

The surge in drug trafficking changed the relationship between DTOs and the state, with traffickers paying taxes directly to politicians and officials (Hernández, 2014, p. 69). However, deviations from implicit pacts were severely prosecuted. For instance, in 1985, the assassination of U.S. Drug Enforcement Administration agent Enrique Camarena led to the arrest of Guadalajara Cartel leaders Ernesto Fonseca Carrillo and Rafael Caro Quintero. Although involved, Miguel Ángel Félix Gallardo avoided arrest until 1989. The division of the Guadalajara Cartel into factions sparked violent conflicts, prompting criminal groups to form private militias for territorial defense and attacks (Trejo and Ley, 2017).

The PRI's dominance weakened after the Partido Acción Nacional (PAN) won the 2000 presidential election, following earlier opposition victories at the municipal and gubernatorial levels. This political shift disrupted coordination among local, federal government, and security forces, leading criminal organizations to switch from bribery to coercive violence and contest territory. In 2006, PAN candidate Felipe Calderón narrowly defeated Andrés Manuel López Obrador (AMLO). Despite increasing drug-related violence in some routes, Calderón inherited one of the safer countries in Latin America, with a national homicide rate at an all-time low of 10 murders per 100,000 inhabitants (Wainwright, 2017). To gain public trust, Calderón declared war on drug trafficking with $2.5 billion in military aid. His administration, marked by heavy militarization, saw the arrest or killing of 25 drug lords and 160 lieutenants in six years—double the captures of the previous administrations (Calderón et al., 2015). His term ended in 2012 with a doubled homicide rate (Fig. 1) and rumors that Secretary of Public Security Genaro García Luna had aided the Sinaloa cartel. García was later arrested and convicted in 2023. The targeted crackdowns led to the increased militarization of drug traffickers, with groups like the Zetas, composed of former soldiers, joining the Gulf Cartel.

The escalation of homicide rates during Calderón's six-year administration resulted from conflicts between and within drug cartels, which expanded criminal activity into new areas and led to the proliferation of factional splits within organizations after leaders were captured (Durán-Martínez, 2017). While leadership decapitation briefly decreased violence, it surged again as sub-leaders competed for control, causing organizational fragmentation (Phillips, 2015). Areas with close elections of PAN mayors, more aligned with President Calderón's crackdowns witnessed a greater increase in violence (Dell, 2015). The increased cocaine seizures in Colombia during this period contributed to elevated drug prices. The more profitable market led to a 10–14% spike in homicides in Mexico, particularly along smuggling routes to the United States (Castillo et al., 2020).

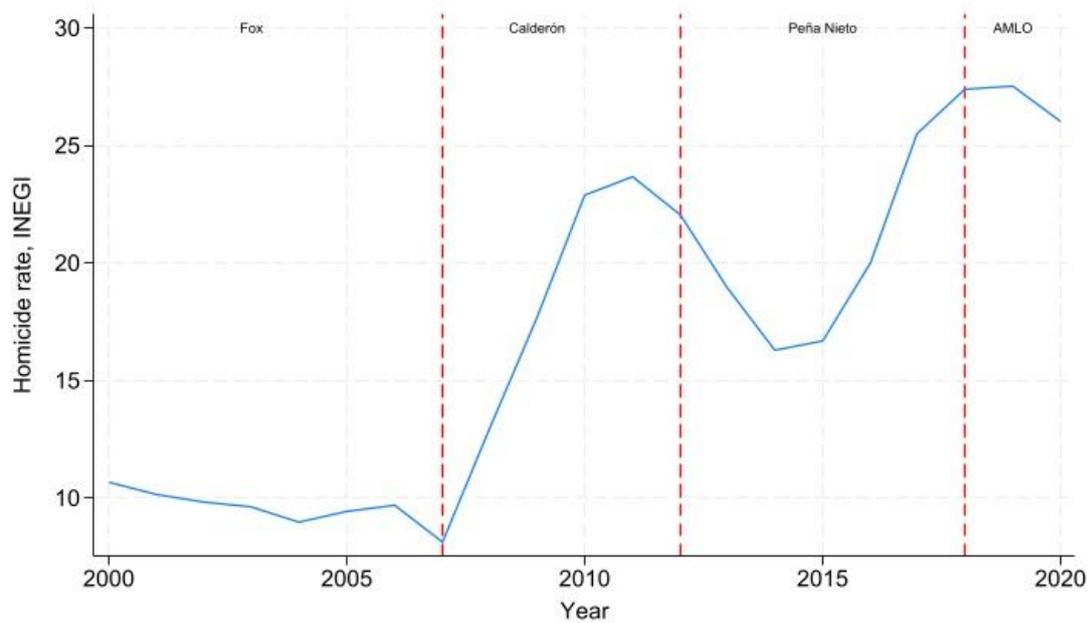

**Fig. 1.** Annual homicide rate in Mexico

The conflict among DTOs in Mexico has persisted despite a decline in arrests and killings of drug kingpins (Fig. 2). Facing challenges in smuggling narcotics into the United States, DTOs have fragmented and ventured into alternative criminal markets. By 2010, major DTOs like the Zetas and Gulf cartel turned their attention to oil theft, known as *huachicol*. This involves bribing workers from Petróleos Mexicanos (Pemex) to obtain pipeline schedules, extracting gasoline and diesel from pipelines, and bribing local governments and police to avoid detection. About 10% of stolen fuel is sold directly to the public, while 90% is sold to petrol stations and large businesses in nearby areas where DTOs have secured local protection (Stevenson, 2017). Some DTOs also exploit their smuggling routes to export substantial quantities of stolen fuel to the United States (Seth, 2018).

Pemex, one of the world's largest oil companies representing nearly 20% of the government's budget, funded the government's development programs by maintaining low taxes and selling gasoline at subsidized prices (Seth, 2018). However, the company's productivity declined due to aging infrastructure and dwindling exploration investments from the mid-2000s. To meet domestic demand, Mexico increasingly relied on imported gasoline. The Calderón administration responded by incrementally raising fuel prices to consumers, leading to heightened violence in the oil theft market as drug trafficking organizations competed for the more lucrative market. The subsequent administration of Peña Nieto introduced energy reforms, causing fuel prices to rise by up to 20% in 2017, leading to fuel price deregulation a year later (International Crisis Group, 2022).

The shift in criminal focus has led to a surge in violence in regions previously less affected by drug trafficking. A notorious case is Puebla, a once-peaceful state where multiple gasoline pipelines intersect. It has witnessed a surge in violence as drug traffickers entered the oil theft business. Existing oil thieves initially paid tributes to DTOs for operational freedom but faced brutal reprisals by the Zetas over turf disputes (International Crisis Group, 2022). Oil theft, now as profitable as drug trafficking, has become a focal point for these criminal groups (Seth, 2018). In 2017, the Cartel Jalisco Nueva Generación (CJNG) was attracted to rising gasoline prices and engaged in oil theft in Puebla, killing hundreds of oil thieves not linked to their organization (Seth, 2018). The CJNG then repeated the purge in Querétaro, Hidalgo, and Guanajuato, clashing with the Santa Rosa de Lima Cartel, which is also dedicated to oil theft in these states. The cycle of retaliation between the two DTOs continues due to their militarized capabilities and infiltration of local governments. Similar disputes over oil theft persist in the states of Jalisco, Michoacán, Oaxaca, Veracruz, and Tamaulipas (Seth, 2018).

In 2018, the then-newly elected Andrés Manuel López Obrador promised to reduce violence in Mexico. As part of this effort, he temporarily shut down refineries, distributed gasoline using tanker trucks instead of the pipeline network, and deployed 4,000 soldiers to strategic locations (Jones & Sullivan, 2019). Although this strategy disrupted gasoline distribution throughout the country and temporarily reduced oil theft, it resumed a few months later and spread to other states with oil pipelines. By 2023, oil theft reached an estimated $5.2 billion despite stronger penalties being imposed (Arzate, 2023). Fig. 2 illustrates the increasing number of clandestine oil taps discovered by the government along the oil pipeline networks. While this number does not indicate the exact amount of oil extracted, it serves as an indicator of the growing prominence of oil theft and the affected states.

Along with the spike in drug-related violence and oil theft, political violence also rose. During the 1980s and 1990s, Mexico experienced occasional political killings and periodic conflicts between PRI members and regional opposition groups (Calderón Molgóra, 1994). One notable isolated incident was the killing of the PRI presidential candidate and likely winner, Luis Donaldo Colosio, before the 1994 elections. This crime took place in Tijuana, a city infamous for drug trafficking, and today the motives of the assassination remain unclear. Since then, political violence has been far more frequent but has predominantly affected mayors.

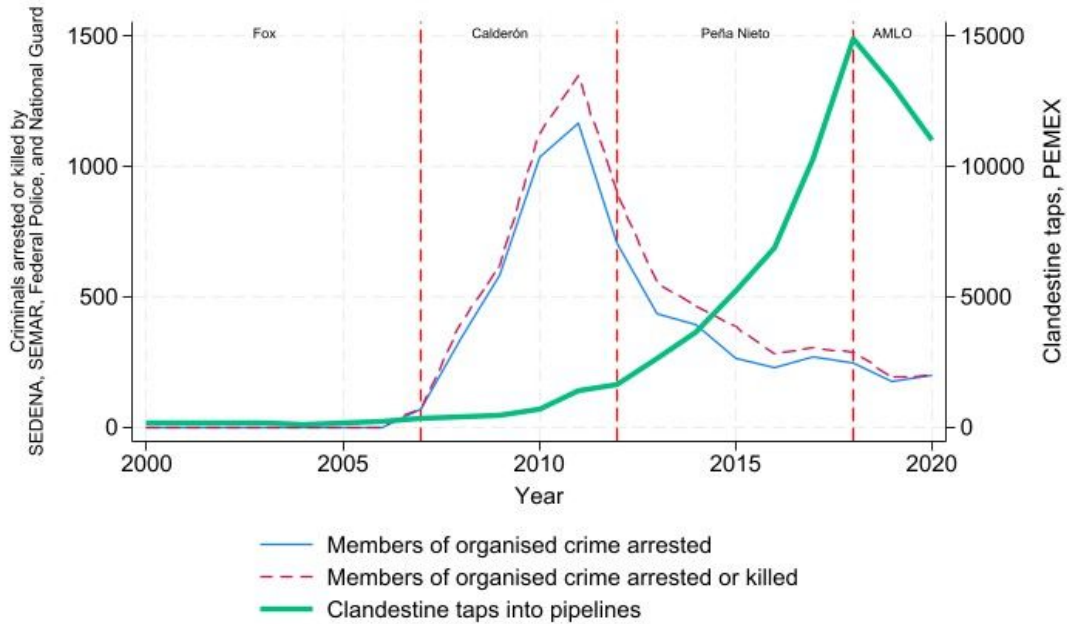

**Fig. 2.** Arrests of members of organized crime and oil theft

To analyze political assassinations, we use PAIAMEX. This dataset was gathered through a manual review of government reports, local media, and historical news archives. To minimize bias, we cross-referenced multiple sources and consulted earlier studies like Esparza & De Paz Mancera (2018) and Trejo & Ley (2021), focusing on earlier periods.

PAIAMEX recorded nearly 500 political assassinations during 2000–21, primarily affecting incumbent mayors, followed by former mayors and candidates for various municipal roles (Fig. 3). Since the mid-2000s, the number of politicians assassinated has steadily increased, reaching an all-time high in 2018. Most of the victims were incumbent mayors and former mayors (Fig. 4). During the onset of the drug war, the main targets of political assassinations were members of opposition parties, possibly because they had less security assistance than the incumbent party, PAN, at the time. The pattern of political assassinations has shifted in recent years. Both opposition and incumbent parties have been targeted, likely due to the ongoing increase in political alternation and the ongoing turf war (Fig. 5).

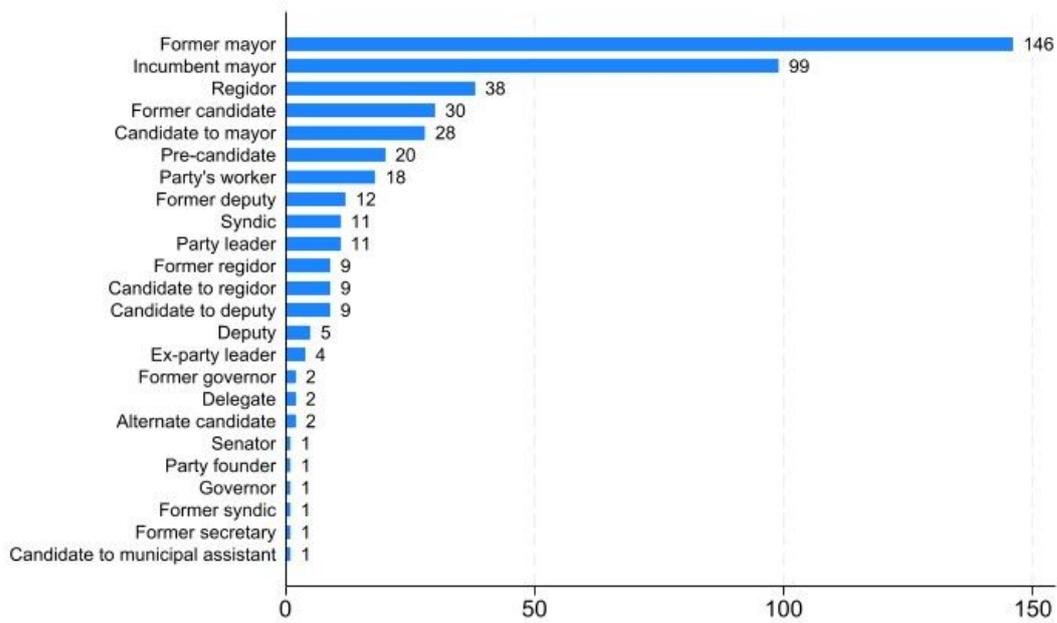

**Fig. 3.** Political assassinations in Mexico 2000–21, PAIAMEX

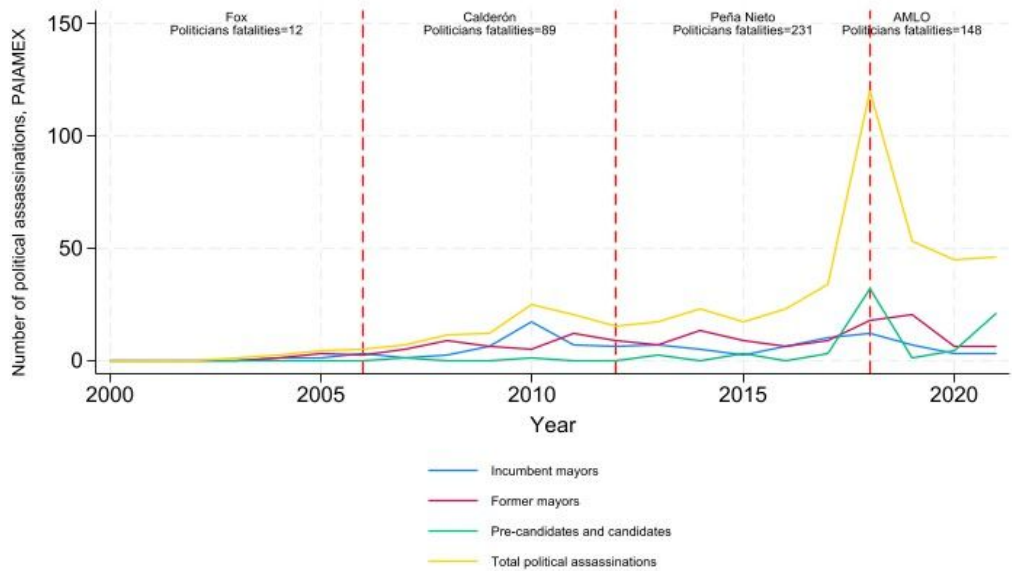

**Fig. 4.** Political assassinations over time

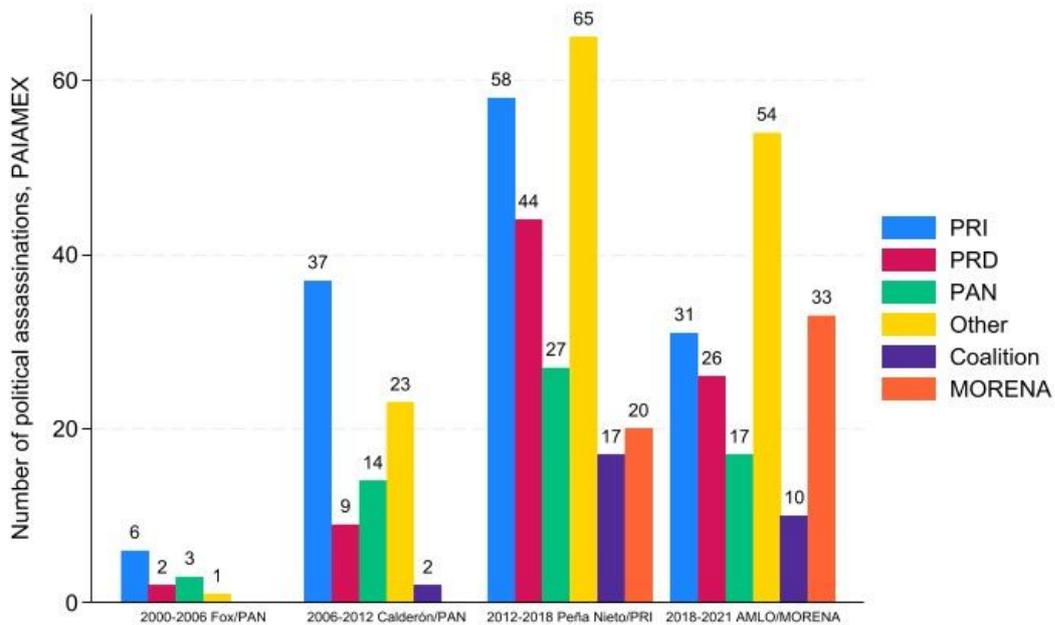

**Fig. 5.** Political affiliation of assassinated politicians

Since political killings are committed with flagrant impunity, it is hard to determine the real intentions behind them. Media information compiled by PAIAMEX reveals a striking pattern: 82% of these political crimes were perpetrated by people wielding high-caliber weapons, presumed members of organized crime (Fig. 6). In only 4% of instances is the presence of a lone assassin noted. However, ties to organized crime cannot be ruled out, and neither can the nature of the potential political motive.

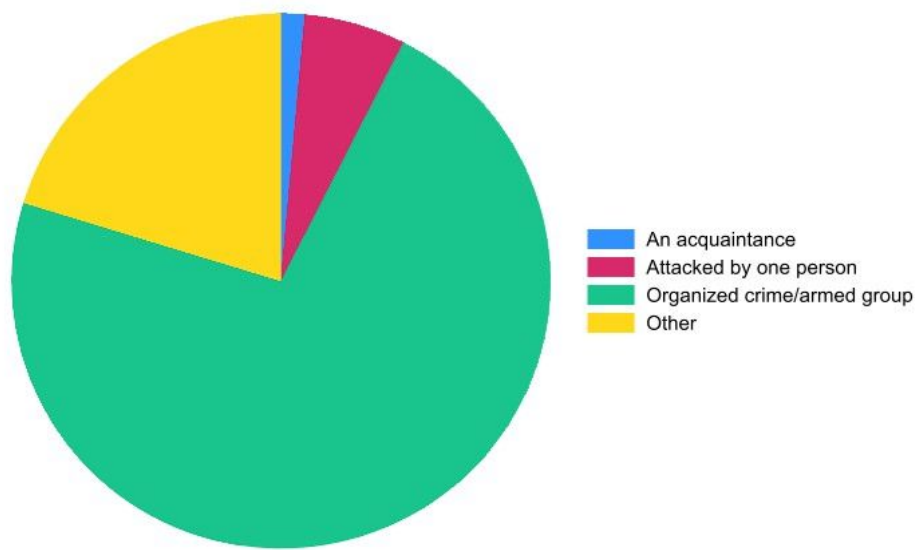

**Fig. 6.** Assassins of politicians in Mexico, 2000–21

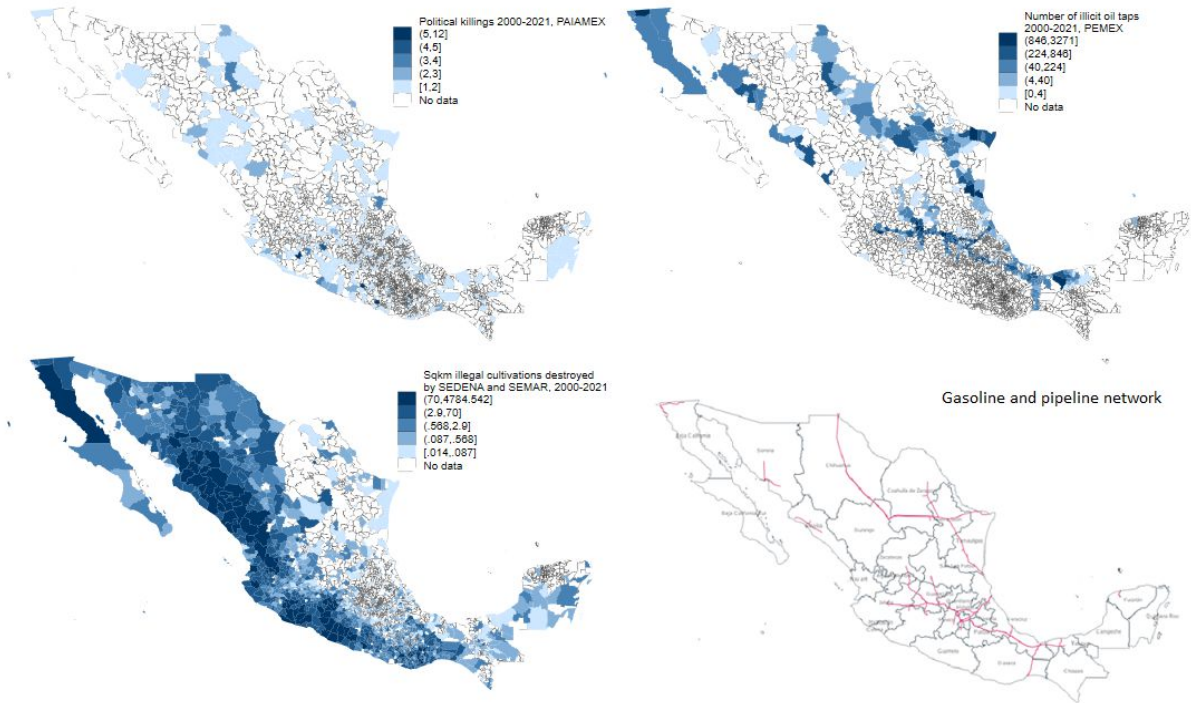

**Fig. 7.** Political assassinations, oil theft, and destruction of illegal cultivates, 2000–21

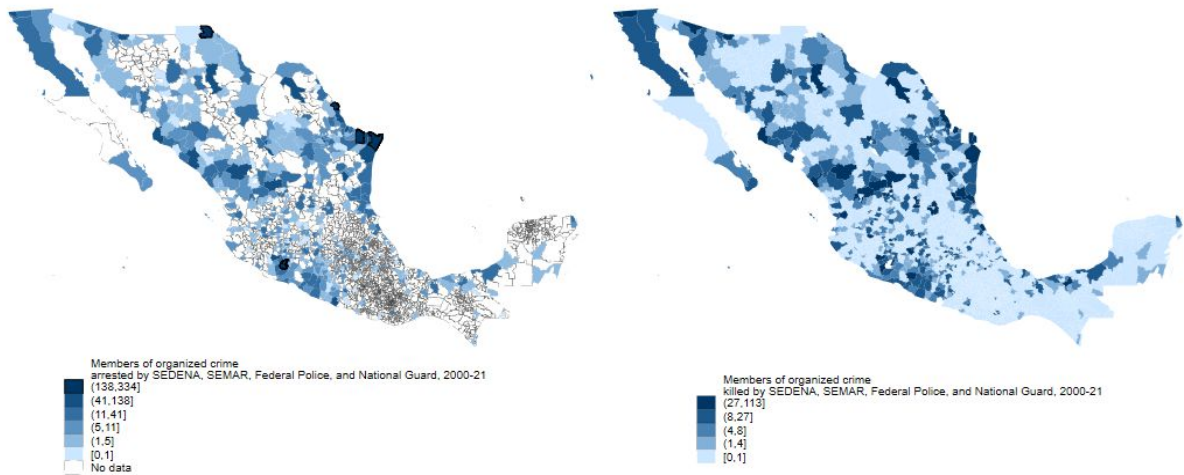

**Fig. 8.** Members of organized crime arrested or killed by the state, 2000–2

An alternative approach to demonstrating the involvement of organized crime in political assassinations is to examine the geographical distribution of these killings during 2000–21. Fig. 7 reveals a strong spatial correlation between political assassinations, the oil pipeline network, and the locations of clandestine oil taps on both the east and west coastlines. Particularly notable is the western region, which shows a significant overlap among political assassinations, oil theft, the destruction of illicit crops, and the arrests or killings of organized crime by all the various security forces during this period (Fig. 8). The top ten states with the highest number of political assassinations include Guerrero, where DTOs contest territories for opium cultivation and smuggling. Other states where drug trafficking and oil theft intersect, such as Oaxaca, Veracruz, Michoacán, Puebla, Estado de México, Chihuahua, Guanajuato, Jalisco, and Tamaulipas, also feature prominently. Together, these states account for nearly 75% of all political assassinations in the country during 2000–21 (Fig. A1).

To sum up, it is clear that in response to the war on drugs, criminal organizations have shifted to new activities, with oil theft emerging as one of the most lucrative. We hypothesize that these organizations use rent-seeking violence to capture local governments to extract resources such as gasoline and diesel. These criminal groups also retaliate against local governments, not voters, when facing crackdowns from authorities. Fig. 9 summarizes the hypotheses developed, identifying the targets of political assassinations, explaining why areas where oil theft occurs are at a higher risk of assassinations, and detailing other relevant covariates used to test each hypothesis.

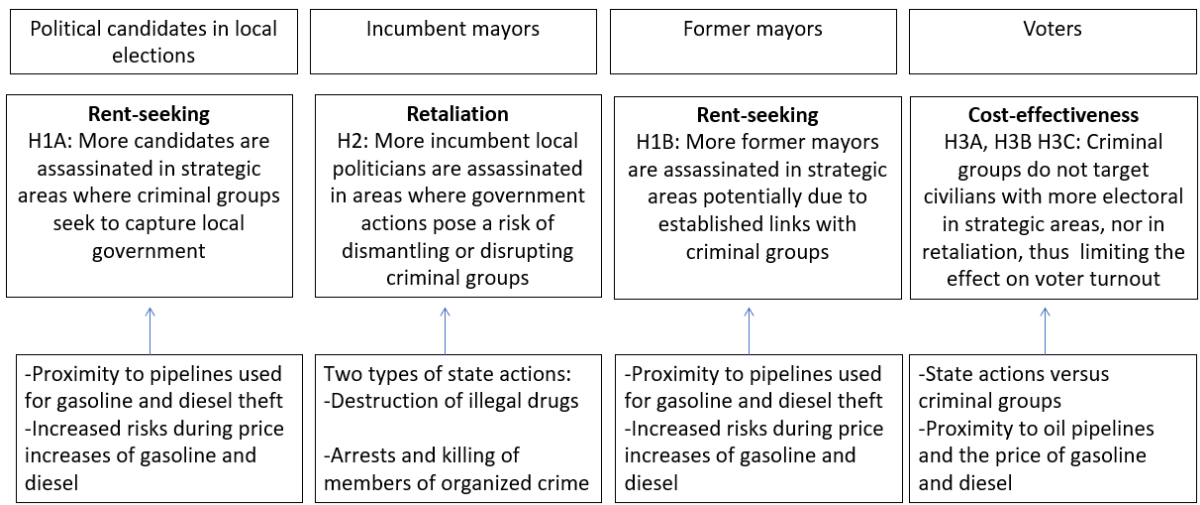

**Fig. 9.** Hypotheses

## 4. Data

To test the H1 (rent-seeking) and H2 (retaliation) hypotheses, we use the number of pre-candidates, candidates, incumbent mayors, and former mayors assassinated from the PAIAMEX dataset. We aggregate this information at the municipality and monthly levels, spanning January 2000 to June 2021, until the latest general election.

For the analysis of the H3 (cost-effectiveness) hypothesis, we use data from the Armed Conflict Location and Event Data Project (ACLED) for 2018–21, as earlier data are not available. From ACLED, we draw the number of fatalities from electoral violence attacks targeted at civilians, excluding deaths or attacks directed at politicians. Violence against civilians is defined as an armed or violent group attacking unarmed civilians who are not engaged in political violence (Raleigh & Dowd, 2016). We also analyze voter turnout and whether the incumbent party emerged victorious in mayoral elections during 2000–21, sourced from each of the 32 electoral institutes nationwide.

Due to national security concerns, the government does not disclose municipality-specific information on the extent of oil theft. To test H1 and H3, we divide the distance in kilometers from the center of each municipality to the nearest oil pipelines by the monthly prices of gasoline and diesel. The oil pipeline infrastructure predates the drug trafficking conflict. Fuel prices are now deregulated and depend on market conditions including international fluctuations, thus also exogenous.

To test H2 and H3, we use a monthly dummy variable, at the municipality level to indicate whether the federal police, army, or navy have destroyed illegal cultivations of marijuana or opium, seized illegal drugs, or dismantled drug labs used to produce synthetic drugs. We chose a dummy variable due to challenges in aggregating data with different units of measurement. We also track the number of arrests and killings of members of organized crime by all security forces at the municipality level. Government crackdowns on criminal groups (e.g., destroying illegal drugs and arresting/killing criminals) can be endogenous. The next section details the instrumental variable approach used to address this.

We also consider the monthly homicide rate per 100,000 people at the municipal level during 2000–21, using data from the Instituto Nacional de Estadística y Geografía (INEGI). To avoid double counting, we deduct the number of political assassinations and organized crime figures killed by authorities from this homicide rate.

To address security coordination issues, we include a dummy variable indicating whether the mayor's political party matches that of both the governor and the president. Others have used similar measures to assess political coordination against organized crime (Blume, 2017; Gutiérrez-Romero, 2016; Rios, 2015; Trejo & Ley, 2021). We also include the municipal government's annual income, which comprises federal transfers and fiscal revenue adjusted for inflation. This variable is released publicly annually. Additionally, we use annual satellite nightlight data at the municipal level as a proxy for wealth, accounting for economic fluctuations and explaining why certain areas, whether wealthier or poorer, may attract more political assassinations.

Table A1 in the Online Supplementary Appendix presents the summary statistics. Table A2 provides more details about all the variables and sources used.

## 5. Method

We use a panel fixed effects Poisson estimator to examine the role of organized crime in political assassinations. Political assassinations are infrequent, non-negative count events, making the fixed effects Poisson estimator the most suitable, and reliable choice.[1] This estimator produces efficient estimates with count data characterized by having a large number of zero values.[2] Additionally, it accommodates any variance-mean relationship, providing robust estimates even in the presence of over-dispersion or under-dispersion of the data (Wooldridge, 1999). Unlike the negative binomial, the fixed effects Poisson estimator is also robust to distributional failures and serial correlation. As our data aligns more appropriately and reliably with the panel fixed effects Poisson estimator, we prefer using this estimator as described in Equation 1.

---

[1] For instance, the Ordinary Least Squares (OLS) estimator is unsuitable in this context because it is less efficient and robust than the Poisson estimator at handling the discrete nature and distributional properties of count data, particularly the large number of zero values in the dependent variable.

[2] The fixed effects Poisson estimator belongs to the broader family of exponential regressions which are easy to interpret and estimate using pseudo-maximum likelihood (PML). These estimators assume that the conditional mean is equal to the exponential of a linear combination of the regressors. Silva and Winkelmann (2024) demonstrate that, asymptotically, PML estimators of exponential regressions retain their interpretability even when the conditional expectation is misspecified, unlike Tobit or zero-inflated models. They also show that the Poisson PML estimator is one of the most reliable choices even under such misspecification. In contrast, the OLS estimator is much more sensitive to potential misspecifications of the conditional expectation and the distribution of the dependent variable and regressors Due to the differences in their assumptions, functional forms, and specificity of the data analyzed, OLS and Poisson estimates cannot be directly compared.

$$E(y_{it}| d_{it-1}, a_{it-1,}, x_{it}, \beta_i) = \beta_i exp \, (d_{it-1}\beta_1 + a_{it-1}\beta_2 + \, x_{it}\beta_3) + \, u_{it}$$

(1)

where $y_{it}$ denotes the dependent variable, which is the number of politicians assassinated in municipality $i$ in month $t$ during 2000–21. We analyze the assassinations of pre-candidates/candidates, incumbent mayors, and former mayors separately. $u_{it}$ denotes the error term.

We analyze whether government actions against organized crime provoke retaliatory violence, manifested in political assassinations. To do so, we include $d_{it-1}$, a dummy variable indicating whether the municipality, in the previous month, destroyed any illegal cultivations, seized illegal drugs from organized crime, or dismantled labs used to produce synthetic drugs. Additionally, we include $a_{it-1}$, which captures the number of arrests or killings of organized crime members at the municipality level during the previous month. To account for a potential delay in organized crime retaliation, we introduce a one-month lag to both variables. This one-month lag prevents reverse causality, where authorities might increase crackdowns in response to political assassinations. This lag also ensures a clearer chronological sequence of events. We do not aggregate the political assassination data annually precisely to capture the immediate effects of government crackdowns on retaliatory violence.[3]

We capture municipality characteristics using the vector $x_{it}$. This vector includes the potential extractive value of the municipality, represented by the distance to the nearest pipeline divided by the monthly average price of gasoline and diesel. We also account for political alignment, noting whether the same party controlled the municipality, state governorship, and presidency during the month analyzed. Additionally, we include the homicide rate in the previous month, lagged to avoid capturing conflicts arising from drug trafficking organization disputes following leader arrests. To prevent double counting, we exclude political assassinations and killings of organized crime members by authorities from this rate. Furthermore, we account for differences in wealth by including municipalities' annual government income and the logarithm of annual nightlight intensity.

---

[3] We have daily data on political assassinations and arrests of members of organized crime. Only in one incident was a member of organized crime arrested on the same day a politician was killed in the same municipality. However, it remains unclear if the arrest is linked to the killing. Since we only have monthly, not daily, data on the destruction of illegal drugs or the killing of members of organized criminal groups, we prefer to use these lagged variables to prevent reverse causality.

We cluster the standard errors at the municipality level. No assumptions are made on the individual fixed effect, $\beta_i$, as these effects are treated as unknown nuisance parameters in the fixed effects Poisson specification. Pseudo-maximum likelihood is used to estimate the relevant parameters by conditioning on the total count of the dependent variable $\sum y_{it}$. For municipalities where the count of the dependent variable $\sum y_{it}$ is equal to zero for the analyzed period, their estimated probability does not contribute to the likelihood function. Thus, municipalities that did not experience a single political assassination during the entire period are automatically removed from the analysis by this estimator and statistical packages because they are not informative about the parameters to be estimated.

### 5.1 Addressing endogeneity

The fixed effects Poisson estimator assumes that, conditional on covariates, the dependent variable is independent over time given exogenous regressors. However, this estimator can still suffer from endogeneity, leading to misspecification and potential bias. Endogeneity may arise if significant unobserved factors influence both the decisions of when and where authorities combat criminal organizations and the incidence of political assassinations in specific areas.

To address this potential endogeneity, we employ the two-stage pseudo-maximum likelihood fixed effects Poisson estimator (Mullahy, 1997). Firstly, we regress the suspected endogenous variables—namely, the destruction of illicit drugs $d_{it-1}$, and the arrests or killings of organized crime members $a_{it-1}$—on the vector of excluded instruments $z_{it-1}$. Given that both of these suspected endogenous variables are rare events, we use fixed effects Poisson regression to estimate these first-stage regressions.

$$d_{it-1} = \varphi_i \, exp(z_{it-1}\varphi_1 \, + \, x_{it}\varphi_2) + e_{1it-1} \qquad (2)$$

$$a_{it-1} = \pi_i \, exp(z_{it-1}\pi_1 \, + \, x_{it}\pi_2) + e_{1it-1} \qquad (3)$$

In the second-stage, the predictions of the suspected endogenous variables, $\hat{d}_{it-1}$, and $\hat{a}_{it-1}$ are used as regressors for the main dependent variable, as shown in Equation 4. Here $v_{it}$ denotes the error term. We estimate the standard errors using bootstrapping and cluster them at the municipality level to account for the sample variation introduced from the two-stage estimation.

$$y_{it} = \delta_i \exp{(\hat{d}_{it-1}\delta_1 + \hat{a}_{it-1}\delta_2 + + x_{it}\delta_3)} + v_{it} \qquad (4)$$

To test for endogeneity, we use a Wu-Hausman-like test, where $y_{it}$ is regressed on the error terms from Equations 2 and 3 and on $d_{it-1}$, $a_{it-1}$, and $x_{it}$. If these error terms are statistically significant, it suggests that unobserved characteristics affect both the dependent variable $y_{it}$ and the suspected endogenous variables. The second-stage parameters of the FE Poisson estimator identify the causal impact of government actions against criminal organizations on the assassination of politicians.

We use three sets of exogenous instruments, denoted by $z_{it-1}$. The first instrument is the ratio of the Chinese population that arrived in Mexico during the 1930s, which varies by state, to the price of corn in the previous quarter in the municipality. Murphy and Rossi (2020) provide evidence linking the current location of drug trafficking organizations in Mexico to areas where Chinese migrants settled in the early 20[th] century. When migration restrictions in the United States were imposed in the 1880s, Chinese migrants diverted to Mexico, bringing poppy seeds and knowledge of opium production. Some members of the Chinese migrant community developed networks to smuggle goods into the United States. Murphy and Rossi (2020) show that areas with historical Chinese populations are more likely to host drug trafficking organizations today. Additionally, the price of corn in the previous quarter serves as an indicator for farmers deciding whether to diversify into illegal drug production (Dube et al., 2016). Thus, for municipalities with historical Chinese immigrants, lower corn prices correlate with higher levels of illegal cultivation and drug activities. Since corn is a major staple and income source for impoverished farmers, the government risks upheaval by destroying illegal cultivations when corn prices are low. Therefore, price fluctuations influence the likelihood of government actions against illegal drug producers and associated organized crime.

The second instrument involves the interaction between municipalities' distance to the nearest port and the average annual retail price of cocaine in the United States. The third one is the interaction between the percentage of mountainous territory in municipalities and the yearly average retail price of heroin in the United States. These instruments help explain which areas are more likely to engage in drug cultivation, smuggle drugs from South America, or distribute drugs to other markets, and when these activities are likely to occur. Higher drug prices increase the likelihood that Mexican criminal organizations will resort to violence to compete in a more profitable market (Castillo et al., 2020). As the drug market becomes more profitable, criminal organizations have greater incentives to compete, and the state has greater incentives to combat these organizations by destroying illegal drugs or arresting their members. The retail prices of these drugs are influenced by various international factors, including actions against cocaine production in Colombia and consumer demand and drug policies in the United States (Gutiérrez-Romero, 2024).

All the instruments meet the exclusion criteria, as these historical migration processes and market prices influence the location and timing of government crackdowns against criminal groups but do not directly impact the likelihood of politicians being assassinated. In the next section, we present robust evidence showing that these exogenous instruments indirectly influence political killings by affecting the timing and locations of government actions against criminal groups. Using these instruments, we gain a clearer understanding of the indirect pathways through which law enforcement actions impact political violence.

## 6. Results

We present the findings on the motives for the killings of Mexican politicians. Table 1 shows the results for candidates and incumbent mayors. Table 2 focuses on former mayors. Each table displays the Poisson fixed effects estimator with and without instrumental variables (IV), featuring two models for each specification. The first model includes all previously mentioned controls. The second model includes all these variables, but the number of members of organized crime arrested excludes the number of criminals killed by law enforcement, as their deaths may have different effects.

The first-stage regression results for the two IV models presented in Table A3 indicate that all instruments are statistically significant and have the expected impacts. Additionally, we present the equivalent F-test, a Wald Chi$^2$ statistic for the excluded instruments, indicating that the instruments are strongly correlated with the endogenous variables. The p-values of the endogeneity tests are shown in the bottom row of Table 1 and Table 2. These suggest evidence of endogeneity for both the assassination of candidates and incumbent mayors. Consequently, for these politicians, the specifications of the fixed effects Poisson estimator with IV (Table 1, columns 5–8) should be preferred. For former mayors, there is no evidence of endogeneity; hence, the specification without IV should be preferred.

**Table 1**

Political assassinations of candidates and incumbent mayors 2000–21, Incidence Rate Ratios.

| | (1) Pre- and candidates | (2) Pre- and candidates | (3) Incumbent Mayor | (4) Incumbent Mayor | (5) Pre- and candidates | (6) Pre- and candidates | (7) Incumbent Mayor | (8) Incumbent Mayor |
|---|---|---|---|---|---|---|---|---|
| | | Fixed Effects Poisson | | | | Second-Stage IV Fixed Effects Poisson | | |
| State's actions against illegal drugs (destroyed cultivations, seized drugs, dismantled labs) in the previous month | 2.650** | 2.644** | 0.768 | 0.765 | 44.141 | 48.037 | 0.739 | 1.208 |
| | (1.145) | (1.145) | (0.315) | (0.315) | (112.548) | (114.934) | (1.623) | (2.502) |
| Number of members of organized crime arrested or killed in the previous month | 1.075 | | 1.056 | | 1.205 | | 2.470*** | |
| | (0.076) | | (0.040) | | (0.584) | | (0.705) | |
| Number of members of organized crime arrested in the previous month | | 1.078 | | 1.152 | | 1.201 | | 2.338*** |
| | | (0.075) | | (0.118) | | (0.545) | | (0.626) |
| Distance to the oilpipeline divided by the average price of gasoline and diesel | 0.231*** | 0.231*** | 0.814*** | 0.815*** | 0.119** | 0.119** | 0.873 | 0.866 |
| | (0.069) | (0.069) | (0.061) | (0.061) | (0.116) | (0.116) | (0.122) | (0.120) |
| Homicide rate in the previous month (excluding criminals and political assassinations) | 1.044*** | 1.044*** | 1.006 | 1.006 | 1.040 | 1.039 | 0.998 | 0.998 |
| | (0.012) | (0.012) | (0.005) | (0.005) | (0.035) | (0.035) | (0.011) | (0.011) |
| Political coordination: The municipality was ruled by the same party as its respective state and presidency | 0.466 | 0.465 | 0.421** | 0.422** | 0.764 | 0.759 | 0.781 | 0.738 |
| | (0.276) | (0.276) | (0.184) | (0.184) | (2.394) | (2.304) | (1.268) | (1.199) |
| Annual income of the government municipal budget, in real terms | 1.000 | 1.000 | 1.000 | 1.000 | 1.000 | 1.000 | 1.000 | 1.000 |
| | (0.000) | (0.000) | (0.000) | (0.000) | (0.000) | (0.000) | (0.000) | (0.000) |
| Annual nightlight in logarithm | 0.695 | 0.695 | 1.175 | 1.171 | 0.376 | 0.366 | 2.413* | 2.067* |
| | (0.297) | (0.297) | (0.312) | (0.310) | (0.297) | (0.275) | (1.100) | (0.877) |
| Observations | 11,756 | 11,756 | 15,766 | 15,766 | 8,907 | 8,907 | 12,453 | 12,453 |
| Wald Chi2 | 97.31 | 98.38 | 18.49 | 18.20 | 38.17 | 38.21 | 35.29 | 35.34 |
| Log pseudolikelihood | -230.7 | -230.7 | -391.9 | -391.8 | -161.4 | -161.4 | -298.5 | -298.4 |
| Endogeneity test | | | | | 4.77 | 4.82 | 23.52 | 22.68 |
| P-value | | | | | 0.029 | 0.090 | 0.000 | 0.000 |

Robust standard errors, clustered at the municipality level, are in parentheses. IV specifications also have bootstrapped standard errors. First-stage regression for columns 5–8 is in Table A3. Significance levels *** p<0.01, ** p<0.05, * p<0.1.

### 6.1 Rent-seeking hypothesis

The regression coefficients are presented as Incidence Rate Ratios (IRRs). An IRR of 1 means no effect, IRR > 1 indicates an increase, and IRR < 1 indicates a decrease in the expected number of political assassinations for a one-unit increase in the covariate analyzed. We also use (IRR coefficient-1)×100 to show the expected percentage change in the number of political assassinations for a unit increase in the covariate.

Table 1 provides support for the H1A hypothesis, which states that criminal groups are more likely to kill candidates in strategic extraction areas. This is evident in municipalities closer to oil pipeline networks utilized by drug traffickers to steal oil. In these municipalities, a unit increase in the ratio of the distance in kilometers from oil pipelines to the average price of gasoline and diesel is associated with an 88% decrease [(0.119-1)*100] in the expected number of assassinations of candidates. Conversely, this indicates that the closer the proximity to the pipelines, the higher the chance of candidates being killed. The goal of criminal groups seems to infiltrate incoming governments, particularly during elections and periods when oil theft becomes more profitable.

**Table 2**

Political assassinations of former mayors 2000–21, Incidence Rate Ratios.

| | (1) Former mayor | (2) Former mayor | (3) Former mayor in politics, business, or an unknown profession | (4) Former mayor in politics, business, or an unknown profession | (5) Former mayor | (6) Former mayor | (7) Former mayor in politics, business, or an unknown profession | (8) Former mayor in politics, business, or an unknown profession |
|---|---|---|---|---|---|---|---|---|
| | Fixed Effects Poisson | | | | Second-Stage IV Fixed Effects Poisson | | | |
| State's actions against illegal drugs (destroyed cultivations, seized drugs, dismantled labs) in the previous month | 0.979 | 0.978 | 0.893 | 0.892 | 13.905 | 11.860 | 10.107 | 8.589 |
| | (0.287) | (0.287) | (0.284) | (0.284) | (23.174) | (18.258) | (18.707) | (14.805) |
| Number of members of organized crime arrested or killed in the previous month | 1.047** | | 1.048** | | 0.746 | | 0.742 | |
| | (0.024) | | (0.024) | | (0.211) | | (0.216) | |
| Number of members of organized crime arrested in the previous month | | 1.053** | | 1.054** | | 0.759 | | 0.756 |
| | | (0.027) | | (0.027) | | (0.201) | | (0.207) |
| Distance to the oilpipeline divided by the average price of gasoline and diesel | 0.752*** | 0.752*** | 0.747*** | 0.747*** | 0.614*** | 0.615*** | 0.611*** | 0.613*** |
| | (0.053) | (0.053) | (0.053) | (0.053) | (0.092) | (0.091) | (0.100) | (0.099) |
| Homicide rate in the previous month (excluding criminals and political assassinations) | 1.005** | 1.005** | 1.002 | 1.002 | 1.001 | 1.001 | 0.999 | 0.999 |
| | (0.002) | (0.002) | (0.003) | (0.003) | (0.007) | (0.007) | (0.008) | (0.008) |
| Political coordination: The municipality was ruled by the same party as its respective state and presidency | 0.710 | 0.711 | 0.733 | 0.734 | 0.595 | 0.606 | 0.587 | 0.598 |
| | (0.257) | (0.257) | (0.295) | (0.296) | (0.309) | (0.311) | (0.367) | (0.370) |
| Annual income of the government municipal budget, in real terms | 1.000** | 1.000** | 1.000 | 1.000 | 1.000 | 1.000 | 1.000 | 1.000 |
| | (0.000) | (0.000) | (0.000) | (0.000) | (0.000) | (0.000) | (0.000) | (0.000) |
| Annual nightlight in logarithm | 0.995 | 0.996 | 1.099 | 1.100 | 0.623 | 0.655 | 0.745 | 0.784 |
| | (0.179) | (0.179) | (0.217) | (0.217) | (0.216) | (0.207) | (0.264) | (0.253) |
| Observations | 24,564 | 24,564 | 21,597 | 21,597 | 21,597 | 16,892 | 14,227 | 14,227 |
| chi2 | 45.00 | 45.11 | 31.90 | 31.97 | 33.49 | 33.48 | 20.56 | 20.56 |
| Log pseudolikelihood | -608.5 | -608.4 | -530.4 | -530.2 | -424.9 | -424.9 | -358.2 | -358.2 |
| Endogeneity test | | | | | 0.640 | 1.360 | 0.180 | 0.870 |
| P-value | | | | | 0.424 | 0.505 | 0.669 | 0.648 |

Robust standard errors, clustered at the municipality level, are in parentheses. IV specifications also have bootstrapped standard errors. The first-stage regression for columns 3–4 is in Table A3. Significance levels *** p<0.01, ** p<0.05, * p<0.1.

Former mayors, as previously indicated, are the second most killed group of politicians. According to independent intelligence reports, some of the deceased former mayors had business ties with organized crime, demonstrating the dangerous nature of narcocracies (Mejía, 2021). To investigate this further, Table 2 examines assassinated former mayors, regardless of when they left office. We also analyze separately the following sub-group: those who remained in politics at the time of their assassination (15%), entrepreneurs (27%), and those whose work status was unknown at the time of their death (41%).

Table 2 indicates that a unit increase in the ratio of the distance to oil pipelines to the average price of gasoline and diesel results in a 25% reduction [(0.75-1)*100] in the expected number of assassinations of former mayors (Table 2, columns 1–4). In other words, we find support for H1B, which asserts that former mayors closer to extractive-value areas are more likely to be killed.

*6.2 Retaliation hypothesis*

We now test more directly the involvement of organized crime in these political assassinations. As expected, we find no association between government crackdowns against criminal groups and the assassinations of candidates (Table 1, columns 5–6). After all, these candidates were not yet in charge of local governments to suffer from this kind of retaliation. In contrast, supporting H2, the expected number of assassinations of incumbent mayors increases by a factor of 2.47 for every additional member of organized crime arrested or killed. The risk of only arresting members of organized crime is slightly lower, at 2.33 (Table 1, columns 7–8).

Former mayors, including those who remained in politics, were entrepreneurs, or had an unknown profession at the time of their death, had nearly a 5% increased risk of assassination [(1.05-1)*100] when authorities arrested or killed members of organized crime (Table 2, columns 1–4). This slight association may suggest a likely connection between these politicians and criminals, which may have led to their demise.

It is also possible that these former mayors were targeted for actions taken during their time as mayors. On average, these former mayors had left their post seven years before their assassination, with 25% leaving a decade before, and one outlier leaving 34 years prior. To focus on potential retaliation for actions during their administrations, we re-analyze only those former mayors who left their posts in the previous administration less than four years ago. This involves aggregating data per administration term that ended less than four years ago, instead of monthly. So, we include the total number of members of organized crime who were killed or arrested during the former mayor's administration. We include a dummy variable for whether the government destroyed illegal drugs. While it does not measure the intensity of destruction efforts, it helps control for crackdowns against drug trafficking.

Since we are aggregating data from the administration that ended less than four years ago, we use a slightly revised IV regression. As instruments, we use the interaction between the distance to the nearest port and the retail price of cocaine in the United States. Additionally, we use the interaction between the retail price of heroin and the municipalities' percentage of mountainous territory. The instruments are statistically significant, as shown in the first-stage regression (Table A4).

For mayors who left office in the previous administration, arrests or killings of members of organized crime during their tenure do not increase their likelihood of being killed later (Table A5). However, former mayors who left office three years ago and remained in politics, business, or an unknown profession faced a higher risk of assassination if illicit drugs were destroyed during their tenure (Table A5, columns 3–4).

We also investigate the hypothesis that criminal organizations kill local politicians who refuse to hand over tax income, as suggested by Trejo & Ley (2021). Our findings show no link between municipal government annual income (including fiscal revenue and federal transfers) and political assassinations. This result holds for local candidates, incumbent, and former mayors (Tables 1, 2, and A5). The findings remain robust regardless of whether IV is considered.

Earlier studies in Italy have also used the homicide rate as a proxy for the presence of organized crime, and in Mexico as a proxy for ongoing competition among criminal organizations (Alesina et al., 2019; Hernández Huerta, 2020). These studies show that the homicide rate is associated with political assassinations. To test this hypothesis and prevent double counting, we subtract the number of assassinated politicians and killed members of organized crime from the overall homicide rate. Table 1 shows that the association between the previous month's homicide rates and political assassinations is statistically insignificant (columns 5–8). Similarly, Tables 2 and A5 show no association between the homicide rate and the assassination of former mayors. This suggests that other factors, like retaliation for government crackdowns, motivate the assassinations of mayors.

We also investigate whether certain political parties are at a greater risk of suffering political assassinations. To test this, we include a dummy indicating whether the local incumbent belongs to the same party as the governor and the president. Our findings show that this type of political coordination is statistically insignificant. In other words, the risk of assassination for candidates, incumbent mayors, or former mayors is not related to whether the local incumbent belongs to the ruling party or the opposition. This is the case regardless of whether IV is used (Table 1, 2, and A5).

Earlier, we mentioned that political decentralization in Mexico led to an increase in overall violence as authorities found it harder to coordinate efforts to tackle criminal groups (Dell, 2015; Gutiérrez-Romero, 2016; Rios, 2015). However, our results suggest that this type of political decentralization is not associated with the recent rise in the killing of politicians. As shown earlier in Fig. 5, over the 2000–21 period, members of all political parties have suffered assassinations, regardless of whether they belonged to the incumbent presidential government or not.[4]

### 6.3 Political violence targeted towards voters and electoral outcomes

An important aspect that has received little attention in the literature is how political violence produced by organized crime in Mexico has affected regular voters. Unfortunately, no single database of electoral violence exists for our entire analysis period, 2000–21. However, ACLED has been systematically collecting data on electoral violence in Mexico since 2018. Since then, political assassinations in the country have skyrocketed, making it an ideal period for our analysis.

Electoral violence can manifest in various non-violent ways that are difficult to identify if one only relies on media reporting, as ACLED does. Therefore, we focus on the number of fatalities of civilians derived from electoral violence, which may be more reliably detected from media sources. From these fatalities, we subtract any fatalities of politicians also documented in ACLED.

Our main objective is to examine whether government actions against criminal organizations lead to retaliatory violence against civilians. This analysis may face bias due to potential unobserved factors that may influence both electoral violence and government actions. To test and correct for endogeneity, we combine fixed effects Poisson with IV. We use the same controls as before, denoted by vector $x_{it}$, and the same main IV specification used in our main analysis. The first-stage regression is the same as the one shown in Table A3 (Columns 1–4). Table 3 shows the fixed effects Poisson with and without IV. The bottom row of this table indicates suggestive evidence of endogeneity. Therefore, the panel fixed effects Poisson specifications with IV should be used.

---

[4] We tested alternative specifications, including dummies for coalition-led municipal governments and separate dummies for whether the municipal government was aligned with the governor's or president's party. None of these dummies were statistically significant and therefore not shown in the results.

## 6.4 Cost-effectiveness hypothesis

Following the H3A hypothesis, there is no statistically significant association between electoral violence against civilians and the ratio of the proximity of municipalities to the oil pipeline network to gasoline and diesel prices (Table 3). In other words, there is no indication that organized crime employs more electoral violence to mobilize voters in areas closer to oil pipeline networks than in those further away. Similarly, we find no connection between the income of municipal governments and civilian fatalities due to electoral violence (IRR=1).

**Table 3**

Electoral violence against civilians 2018–21, Incidence Rate Ratios.

| | (1) Fatalities of civilians | (2) Fatalities of civilians | (3) Fatalities of civilians | (4) Fatalities of civilians |
|---|---|---|---|---|
| | Fixed Effects Poisson | | Second-Stage Fixed Effects Poisson IV | |
| State's actions against illegal drugs (destroyed cultivations, seized drugs, dismantled labs) in the previous month | 0.962 | 0.962 | 0.429* | 0.433** |
| | (0.050) | (0.050) | (0.194) | (0.178) |
| Number of members of organized crime arrested or killed in the previous month | 1.031 | | 1.021 | |
| | (0.040) | | (0.120) | |
| Number of members of organized crime arrested in the previous month | | 1.037 | | 1.022 |
| | | (0.040) | | (0.112) |
| Distance to the oilpipeline divided by the average price of gasoline and diesel | 0.923 | 0.922 | 1.046 | 1.046 |
| | (0.049) | (0.049) | (0.070) | (0.070) |
| Homicide rate in the previous month (excluding criminals and political assassinations) | 1.007*** | 1.007*** | 1.010 | 1.010 |
| | (0.002) | (0.002) | (0.007) | (0.007) |
| Political coordination: The municipality was ruled by the same party as its respective state and presidency | 0.825 | 0.826 | 0.884 | 0.884 |
| | (0.104) | (0.104) | (0.101) | (0.098) |
| Annual income of the government municipal budget, in real terms | 1.000 | 1.000 | 1.000** | 1.000** |
| | (0.000) | (0.000) | (0.000) | (0.000) |
| Annual nightlight in logarithm | 1.612* | 1.612* | 3.101*** | 3.100*** |
| | (0.428) | (0.428) | (1.301) | (1.259) |
| Observations | 32,604 | 32,604 | 25,038 | 25,038 |
| Wald Chi2 | 19.32 | 19.50 | 16.11 | 16.13 |
| Log pseudolikelihood | -23753 | -23753 | -18401 | -18401 |
| Endogeneity test | | | 5.350 | 4.560 |
| P-value | | | 0.069 | 0.103 |

Robust standard errors, clustered at the municipality level, are in parentheses. IV specifications also have bootstrapped standard errors. First-stage regression for columns 3–4 is in Table A3. Significance levels *** p<0.01, ** p<0.05, * p<0.1.

Supporting the H3B hypothesis, there is no statistically significant evidence that criminal organizations retaliate against civilians in response to the government destroying illegal drugs or arresting or killing members of organized crime. On the contrary, the expected number of fatalities related to electoral violence is reduced in areas where the government destroys illicit drugs.

Recognizing that various factors influence electoral violence, we investigate whether opposition parties or incumbent parties might face different risks of electoral violence. We find that municipalities governed by the same political party as the respective state and the president do not have a different risk of suffering electoral violence compared to those where this is not the case (Table 3, columns 3–4). Additionally, the expected number of fatalities from electoral violence is not associated with the overall homicide rate (excluding killings of politicians and criminals killed by authorities).

We now focus on the potential effects of political assassinations on electoral outcomes. We focus on voter turnout and the likelihood that the mayor's incumbent party is re-elected.[5] We aggregate the characteristics of the municipalities by electoral term to analyze these aspects. We use panel fixed effects specifications instead of Poisson because electoral outcomes are not rare events.

$$\mu_{ie} = \rho_{1i} + \rho_2 x_{ie} + v_{ie} \tag{5}$$

where $\mu_{ie}$ is our dependent variable. It separately measures two variables of interest. That is, voter turnout (which can theoretically range from 0 up to 100%) and whether the ruling party was re-elected in the municipality (which takes the value of zero or one) $i$ and electoral term $e$. $\rho_{1i}$ is the unobserved time-invariant individual effect and $v_{ie}$ is the error term. The standard errors are clustered at the municipality level.

Our municipality controls, represented by the vector $x_{ie}$ are aggregated for the local administration term, typically of three years, that immediately preceded the election day. We run two models. Both models include controls for the proximity to pipelines divided by the average price of gasoline and diesel; the average municipal income in real terms; the average nightlight in logarithm; and the average homicide rate. In the first model, we also add the number of incumbent mayors, pre-candidates, and candidates who were assassinated during the administration before the election day. In the second model, we instead add only the number of pre-candidates and candidates assassinated during the administration before the election day.

Once again, we may face endogeneity due to unobserved characteristics that may influence both electoral outcomes and government actions against criminal organizations. We address this concern by using panel fixed effects combined with IV. Table A6 presents the first-stage regression, instrumenting whether the government took any action to destroy illegal drugs, and the number of organized crime members killed or arrested during the administration before election day. Due to aggregated covariates at the municipal level per electoral cycle, we adjust our IV specification.

---

[5] Mayors in some states are allowed to be re-elected for consecutive terms. Unfortunately, it is not possible to determine whether the same mayor ran for re-election because the available election data do not include the names of contending candidates.

We use two instruments: the interaction between the distance to the municipality's nearest port and the price of cocaine in the United States, and the interaction between the retail price of heroin in the United States and the municipality's percentage of mountainous territory. The intuition is the same as before: exogenous increases in heroin prices affect Mexican drug profitability and influence government actions, particularly in geospatially relevant municipalities. Given that these state actions are rare events, fixed effects Poisson is used for the first-stage IV regression.

Table A6 shows that all instruments are statistically significant and highly correlated with the instrumented variables. The second-stage, panel fixed effects, with and without IV, are shown in Table A7. The bottom row shows evidence of endogeneity. Thus, the IV specifications should be preferred. These fixed effects IV coefficients are illustrated in Fig.10 as dots, and the 95 percent confidence intervals as lines.

Fig.10 illustrates that voter turnout is unaffected by political assassinations of mayors, pre-candidates, and candidates in the municipality during the electoral cycle before election day, in line with H3C. Fig.10 also shows that the assassination of these politicians does not influence the probability of the incumbent party's re-election. This finding reinforces the argument that retaliatory violence is directed at politicians, not voters, and therefore has little influence on voter turnout and the likelihood of incumbent party re-election.

We also examine the impact of other forms of violence, proxied by homicide rates. To prevent double counting, as before, we subtract the number of political assassinations and criminals killed by authorities from the overall homicide rate. Despite the country's elevated levels of violence, voter turnout falls by at most 0.019 points for a one-point increase in the homicide rate.

Fig.10 demonstrates a notable increase in voter turnout—by approximately 6.6 points—in municipalities where authorities acted against illegal drugs during the election cycle preceding election day. Moreover, the destruction of illicit drugs during this period significantly boosts the incumbent party's re-election probability by 40 percentage points. Additionally, voter turnout rises by 1.8 points for each additional member of organized crime arrested or killed during the administration. These findings indicate that voters favor state actions against criminal organizations, as evidenced by higher turnout rates and a greater likelihood of re-elected parties that engage in such measures.

While the increase in turnout in these areas could be influenced by factors such as electoral malpractice, our analysis suggests that this is unlikely. We also examined voter turnout in municipalities with illicit poppy cultivation using confidential satellite data from the UNODC and the Government of Mexico, covering $10 \times 10km^2$ sample locations between 2014 and 2019 (UNODC and GOM 2021).[6] Our analysis reveals that municipalities with indications of opium cultivation have a slightly lower turnout, by about 1.4 points, compared to areas without such cultivation. These findings imply that crackdowns against criminal organizations can indeed increase voter turnout, and this increase is unlikely to be the result of ballot-stuffing in drug trafficking areas.

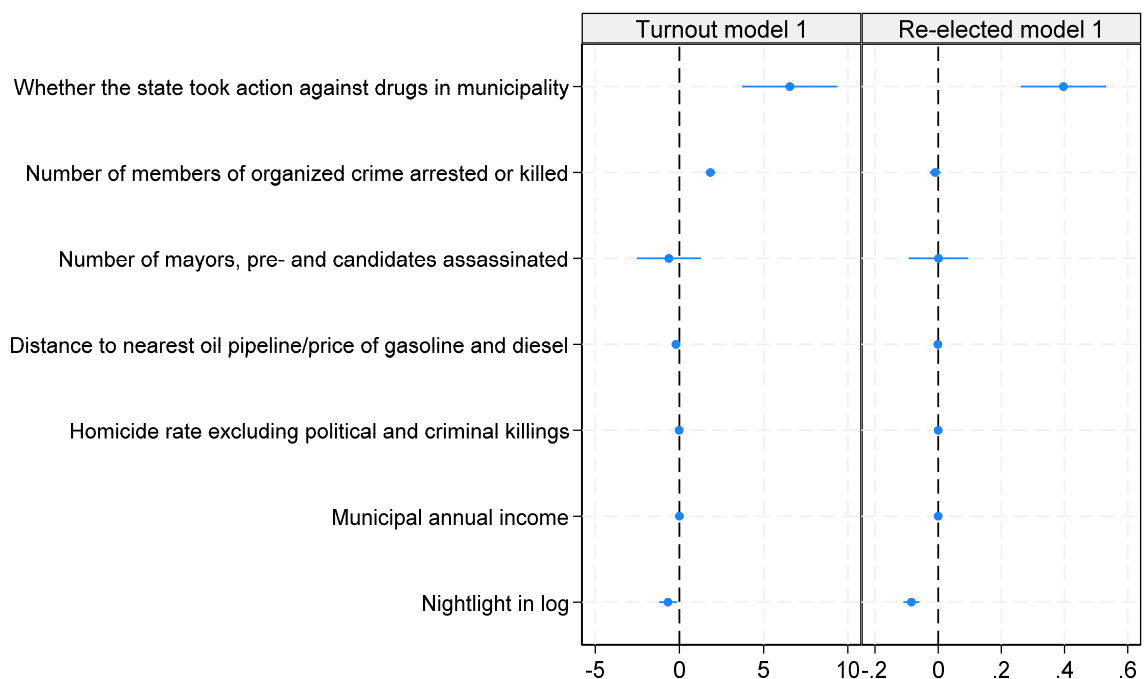

**Fig. 10.** Political violence, voter turnout, and the probability of the incumbent local party winning in mayoral elections, 2000–21. Based on panel fixed effects IV coefficients shown in Table A7

---

[6] We created a dummy variable to identify municipalities with poppy cultivation in the sampled locations monitored. Since this information is unavailable outside the sampled locations, any interpolation might not accurately reflect the actual density of poppy cultivation in non-sampled areas. Due to the confidential nature of the data, we do not present these results in Table A7.

**7. Conclusion**

We analyzed the drivers behind the rise in assassinations of politicians in Mexico during 2000–21 and their impacts on voting behavior. Our theoretical framework identified and tested the areas, targets, and timing of these assassinations. We found that local politicians at the municipal level are the primary targets, with both opposition and incumbent parties facing similar risks. Targeting is primarily driven by organized crime seeking to capture local government for rent-seeking and in retaliation for government crackdowns. Our findings offer three key contributions.

First, in line with our rent-seeking hypothesis, political candidates are more likely to be assassinated in areas of strategic extractive value for criminal organizations. This is particularly true for oil theft, which now provides drug traffickers with a comparable income stream to drug smuggling (Ferri, 2019). Thus, in municipalities closer to the pipelines used for oil theft, the number of political candidates assassinated increases, especially during elections and periods of rising gasoline and diesel prices. Second, our findings indicate that the primary driver behind the increase in the assassination of incumbent mayors is the arrests or killings of organized crime members, which supports our retaliatory violence hypothesis.

Third, criminal organizations find it more cost-effective to assassinate political candidates to capture local governments rather than targeting civilians. This is shown by the lack of increased electoral violence against civilians in municipalities near oil pipelines and the absence of retaliatory violence against voters following crackdowns on organized crime. The targeting of politicians, instead of voters, may explain the lack of impact on voter turnout. Notably, areas where authorities destroy illegal drugs see higher voter turnout and are more likely to re-elect the incumbent local government. These results align with experimental studies showing that exposure to violence increases citizens' support for punishing criminals (García-Ponce et al., 2022).

Our findings are limited to political assassinations and do not cover other strategies such as bribery, threats, non-lethal attacks that candidates, politicians, and their families also face. Nonetheless, our analysis suggests that when criminal groups infiltrate governments, they pose significant threats to peace, economic stability, and democracy by creating entry barriers to politics and compromising policies. These implications extend beyond Mexico to other similar Latin American, African, and Asian countries, where criminal groups kill local politicians with gross impunity. To mitigate these risks, coordinated strategies are needed to restore the rule of law and protect candidates and local politicians, especially in vulnerable areas and during critical times in the electoral process.

**FUNDING**

We acknowledge funding from UNU-WIDER and the Global Challenges Research Fund (GCRF) [RE-CL-2021-01].

**ACKNOWLEDGEMENTS**

We thank the UNODC for providing us with access to data on areas with the presence of illegal crops. We are grateful to Constantino Carreto, David Aban Tamayo, Tania Rodríguez, and Yunuen Rodríguez for research assistance at Simetría during the initial stages of this research. We also thank Ursula Daxecker, Helden DePaz-Mancera, Fabio Ellger, Subhasish Chowdhur, Omar Garcia-Ponce, Kai A. Konrad, Krzysztof Krakowski, Martin Ottman, Anne Pitche, Manuel Pérez Aguirre, David Pérez-Esparza, and Manuel Vélez for their feedback.

# Supplementary Appendix

**Contents**

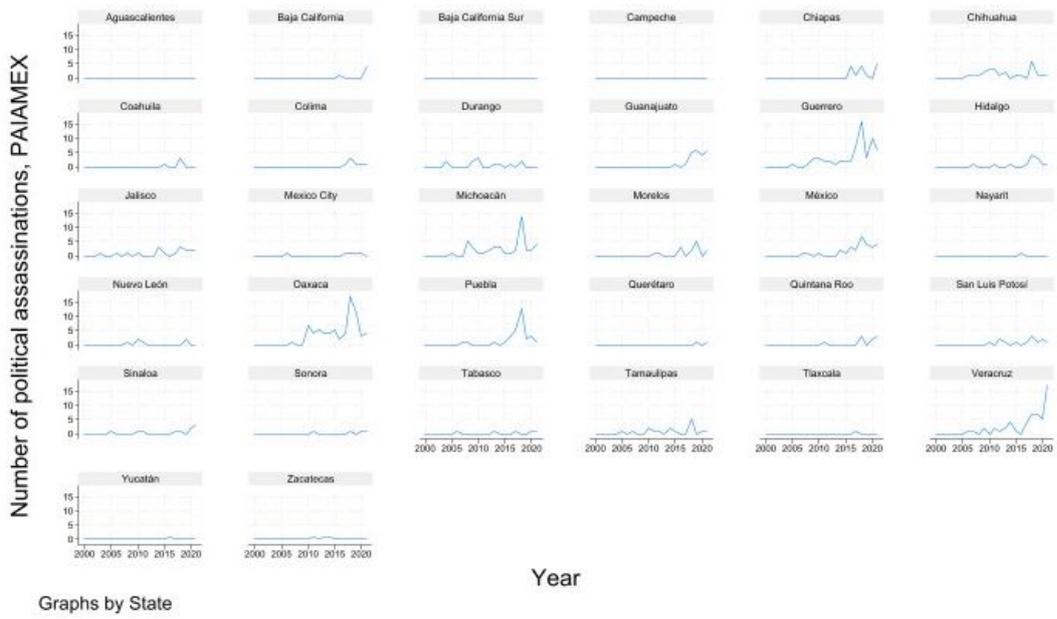

**Fig. A1.** Political assassinations by state, 2000–21

**Table A1.**

Summary statistics by type of political violence experienced.

| | Experienced the killing of a pre-candidate or candidate | | | | | | Experienced the killing of the incumbent mayor | | | | | | Experienced the killing of a former mayor | | | | |
|---|---|---|---|---|---|---|---|---|---|---|---|---|---|---|---|---|---|
| | Months | Obs. | Municipalities | Percent | Mean | Std. dev. | Obs. | Municipalities | Months | Percent | Mean | Std. dev. | Obs. | Municipalities | Percent | Mean | Std. dev. |
| Had monthly state actions against illegal drugs (destroyed cultivations, seized drugs, dismantled labs) | 240 | 11,756 | 49 | 23.21 | | | 15,766 | 66 | 240 | 17.56 | | | 21,597 | 91 | 20.22 | | |
| Number of monthly members of organized crime arrested or killed | 240 | 11,756 | 49 | | 0.04 | 0.54 | 15,766 | 66 | 240 | | 0.02 | 0.48 | 23,247 | 100 | | 0.05 | 0.86 |
| Number of monthly members of organized crime arrested | 240 | 11,756 | 49 | | 0.04 | 0.52 | 15,766 | 66 | 240 | | 0.02 | 0.29 | 23,247 | 100 | | 0.04 | 0.79 |
| Nearest distance in kilometers to oil pipelines | 240 | 11,756 | 49 | | 90.41 | 77.32 | 15,766 | 66 | 240 | | 94.25 | 74.27 | 23,247 | 100 | | 91.58 | 76.52 |
| Average price of gasoline and diesel, in real terms | 240 | 11,756 | 49 | | 13.80 | 2.98 | 15,766 | 66 | 240 | | 13.87 | 3.02 | 23,247 | 100 | | 13.87 | 3.00 |
| Monthly homicide rate (excluding criminals and political assassinations) | 240 | 11,756 | 49 | | 2.13 | 4.75 | 15,766 | 66 | 240 | | 2.57 | 9.41 | 23,247 | 100 | | 2.56 | 9.94 |
| Annual income of the government municipal budget, in real terms and in million pesos | 240 | 11,756 | 49 | | 483.00 | 937.00 | 15,766 | 66 | 239 | | 158.00 | 374.00 | 23,247 | 100 | | 198.00 | 397.00 |
| Annual nightlight in log | 240 | 11,756 | 49 | | 7.78 | 1.43 | 15,766 | 66 | 240 | | 6.60 | 1.57 | 23,247 | 100 | | 6.92 | 1.58 |
| Political coordination: The municipality was ruled by the same party as its respective state and presidency | 240 | 11,756 | 49 | 10.45 | | | 15,766 | 66 | 240 | 14.49 | | | 23,247 | 100 | 12.27 | | |
| Chinese population in the area in 1930 | 240 | 11,756 | 49 | | 205.45 | 371.51 | 15,766 | 66 | 240 | | 142.93 | 212.67 | 23,247 | 100 | | 194.93 | 340.31 |
| Average price of corn in the previous quarter in pesos per ton | 240 | 11,756 | 49 | | 2991.42 | 1082.89 | 15,766 | 66 | 240 | | 3020.99 | 1066.00 | 23,247 | 100 | | 2974.12 | 1077.83 |
| Annual street prices of cocaine, adjusted for purity, per 0.001 gram in the United States | 240 | 11,756 | 49 | | 0.16 | 0.04 | 15,766 | 66 | 240 | | 0.16 | 0.04 | 23,247 | 100 | | 0.16 | 0.04 |
| Street price of heroin adjusted for purity and inflation per gram in the United States | 240 | 11,756 | 49 | | 990.35 | 133.94 | 15,766 | 66 | 240 | | 990.79 | 133.67 | 23,247 | 100 | | 989.25 | 133.24 |
| Percentage of mountainous territory | 240 | 11,756 | 49 | | 4.56 | 7.05 | 15,766 | 66 | 240 | | 9.96 | 16.84 | 23,247 | 100 | | 8.73 | 17.52 |
| Distance to the nearest port | 240 | 11,756 | 49 | | 135.19 | 107.46 | 15,766 | 66 | 240 | | 166.40 | 132.02 | 23,247 | 100 | | 145.17 | 112.19 |

The fixed-effects Poisson estimator used to test all hypotheses discards municipalities where the dependent variable is zero across the period, meaning that no political assassinations occurred, as this does not contribute to the estimation of the likelihood function. Therefore, Table A1 provides the summary statistics only for those municipalities that had at least one political assassination of a pre-candidate/candidate, an incumbent mayor, or a former mayor.

## Table A2.

Data Sources.

| Variable | Description | Level | Term | Source |
|---|---|---|---|---|
| Pre-candidates and candidates killed in political violence | Politicians killed in political violence | Municipality | Monthly 2000-2021 | PAIAMEX |
| Incumbent mayor killed in political violence | Politicians killed in political violence | Municipality | Monthly 2000-2021 | PAIAMEX |
| Former mayor killed in political violence | Politicians killed in political violence | Municipality | Monthly 2000-2021 | PAIAMEX |
| Non-fatal acts of intimidation | Number of intimidation aggressions against politicians with no killings | Municipality | Monthly 2000-2021 | PAIAMEX |
| Incidents of electoral violence | Number of events of electoral violence | Municipality | Monthly 2018-2021 | ACLED, 2022 |
| Fatalities of civilians | Number of civilian fatalities | Municipality | Monthly 2018-2021 | ACLED, 2022 |
| Voter turnout | Voter participation rate | Municipality | Yearly 2000-2021 | Electoral office in each of the 32 states |
| Incumbent party gets re-elected | Dummy variable = 1 when the same political party remains in power | Municipality | Yearly 2000-2021 | Electoral office in each of the 32 states |
| Monthly homicide rate (excluding criminals and political assassinations) | Homicide rate excluding political assassinations and members of organized crime that the state killed per 100,000 inhabitants | Municipality | Monthly 2000-2021 | INEGI, CONAPO, SEDENA, SEMAR, the National Guard, the Federal Police, and PAIAMEX |
| Destruction of illegal drug cultivations | Number of square kilometers of illegal crops of marijuana and poppy destroyed by government state forces | Municipality | Monthly 2000-2021 | SEDENA and SEMAR |
| Seized drugs | Total kilograms of seized drugs (cocaine, fentanyl, heroin, methamphetamine, opium, and marijuana) by government state forces | Municipality | Monthly 2000-2021 | SEDENA, SEMAR, the National Guard, and the Federal Police |
| Drug labs seized | Number of drug laboratories confiscated by government state forces | Municipality | Monthly 2000-2021 | SEDENA, SEMAR, the National Guard, and the Federal Police |
| Members of organized crime who were arrested or killed by authorities | Criminals that belong to organized crime as typified by authorities such as drug trafficking, fraud, extortion | Municipality | Monthly 2000-2021 | SEDENA, SEMAR, the National Guard, and the Federal Police |
| Clandestine oil tapping | Number of illegal oil tapings in pipelines. It is a proxy for oil theft but does not give the number of gasoline or diesel stolen | Municipality | Monthly 2000-2021 | PAIAMEX |
| Annual income of the government municipal budget, in real terms | Municipal annual in real terms, including tax revenue and transfers from the federal government | Municipality | Yearly 2000-2021 | INEGI |
| Nightlight | Satellite nightlight | Municipality | Yearly 2000-2020. The first six months of 2021 are taken from the annual values of the previous year | Earth Observation Group, Payne Institute for Public Policy |
| Political coordination | Dummy variable = 1 when the municipality was ruled by the same party as its respective state and presidency | Municipality | Yearly 2000-2021 | Electoral office in each of the 32 states |
| Chinese population in 1930 | Chinese population by area in 1930 | State | Yearly, 1930 | INEGI |
| Percentage of mountainous territory | Percentage of territory with mountains within the municipality | Municipality | Constant | INEGI |
| Average price of gasoline and diesel | Average price of gasoline and diesel | Municipality | Monthly 2000-2021 | INEGI |
| Price of corn | Average price of green corn | Municipality | Quarter 2000-2021 | INEGI |
| Distance to ports | Distance to the nearest port | Municipality | Constant | Own estimates based on INEGI's geospatial files |
| Annual price of cocaine adjusted for purity and inflation in the United States | Retail price in the United States | U.S. | Annual | Office of National Drug Control Policy (ONDCP) and the United Nations Office on Drugs and Crime (UNODC) |
| Price of heroin adjusted for purity and inflation in 2019 in the United States | Retail price in the United States | U.S. | Annual | Office of National Drug Control Policy (ONDCP) and the United Nations Office on Drugs and Crime (UNODC) |

**Table A3.**

First-stage regression of Tables 1, 2, and 3. Panel fixed effects Poisson, Incidence Rate Ratios.

| | (1)<br>State's actions against drugs | (2)<br>Members of organized crime arrested or killed | (3)<br>State's actions against drugs | (4)<br>Members of organized crime arrested or killed |
|---|---|---|---|---|
| | | Model 1 | | Model 2 |
| Log Chinese population in the area in the 1930s divided by the average price of corn in the previous quarter | 0.772*** | 0.424*** | 0.772*** | 0.464** |
| | (0.034) | (0.135) | (0.034) | (0.150) |
| Distance to the nearest port multiplied by the annual price of cocaine in the United States per kilo | 1.012*** | 1.106*** | 1.012*** | 1.105*** |
| | (0.002) | (0.014) | (0.002) | (0.015) |
| Log Percentage of mountainous territory multiplied by the annual price of heroin in the United States | 0.893** | 25.990*** | 0.893** | 34.599*** |
| | (0.045) | (13.579) | (0.045) | (18.697) |
| Distance to the oil pipeline divided by the average price of gasoline and diesel | 1.077*** | 1.075 | 1.077*** | 1.045 |
| | (0.011) | (0.053) | (0.011) | (0.052) |
| Homicide rate in the previous month (excluding criminals and political assassinations) | 1.002*** | 1.005 | 1.002*** | 1.005 |
| | (0.000) | (0.004) | (0.000) | (0.000) |
| Political coordination: The municipality was ruled by the same party as its respective state and presidency | 0.972 | 0.476*** | 0.972 | 0.493** |
| | (0.028) | (0.129) | (0.028) | (0.140) |
| Annual income of the government municipal budget, in real terms | 1.000*** | 1.000 | 1.000*** | 1.000 |
| | (0.000) | (0.000) | (0.000) | (0.000) |
| Annual nightlight in logarithm | 1.218*** | 0.619** | 1.218*** | 0.641** |
| | (0.028) | (0.126) | (0.028) | (0.129) |
| Observations | 301,898 | 90,329 | 301,898 | 85,564 |
| Log pseudolikelihood | -89141 | -17034 | -89141 | -14957 |
| Wald Chi2 | 267.2 | 177.6 | 267.2 | 153.7 |
| Excluded instruments | | | | |
| Log pseudolikelihood | -98662 | -17989 | -98662 | -15775 |
| Wald Chi2 | 22.63 | 122.4 | 22.63 | 116.8 |

Robust standard errors, clustered at the municipality level, are in parentheses.

Significance levels *** p<0.01, ** p<0.05, * p<0.1.

**Table A4.**

First-stage regression of Table A5. Panel fixed effects Poisson, Incidence Rate Ratios.

| | (1) Whether the state had taken any actions against illegal drugs (destroyed cultivations, seized drugs, dismantled labs) in the administration where the | (2) Number of members of organized crime arrested or killed in the administration where the assassination took place | (3) Whether the state had taken any actions against illegal drugs (destroyed cultivations, seized drugs, dismantled labs) in the administration where the | (4) Number of members of organized crime arrested or killed in the administration where the assassination took place |
|---|---|---|---|---|
| | | Model 1 | | Model 2 |
| Distance to the nearest port multiplied by the annual price of cocaine in the United States per kilo | 1.007*** | 1.114*** | 1.007*** | 1.111*** |
| | (0.001) | (0.035) | (0.001) | (0.035) |
| Log Percentage of mountainous territory multiplied by the annual price of heroin in the United States | 0.511*** | 24.957*** | 0.511*** | 34.752*** |
| | (0.044) | (28.796) | (0.044) | (40.927) |
| Distance to the oil pipeline divided by the average price of gasoline and diesel | 1.019*** | 1.072 | 1.019*** | 1.049 |
| | (0.006) | (0.055) | (0.006) | (0.051) |
| Homicide rate in the previous month (excluding criminals and political assassinations) | 1.000*** | 1.006*** | 1.000*** | 1.005*** |
| | (0.000) | (0.001) | (0.000) | (0.001) |
| Political coordination: The municipality was ruled by the same party as its respective state and presidency | 0.953** | 1.138 | 0.953** | 0.989 |
| | (0.021) | (0.240) | (0.021) | (0.217) |
| Annual income of the government municipal budget, in real terms | 1.000*** | 1.000** | 1.000*** | 1.000** |
| | (0.000) | (0.000) | (0.000) | (0.000) |
| Annual nightlight in logarithm | 1.156*** | 0.620** | 1.156*** | 0.607*** |
| | (0.024) | (0.115) | (0.024) | (0.115) |
| Observations | 7,901 | 2,965 | 7,901 | 2,814 |
| Wald Chi2 | 154.3 | 96.24 | 154.3 | 79.49 |
| Log pseudolikelihood | -4841 | -5562 | -4841 | -4923 |
| Excluded instruments | | | | |
| Wald Chi2 | 62.93 | 15.97 | 62.93 | 14.39 |
| Log pseudolikelihood | -5639 | -6309 | -5639 | -5558 |

Robust standard errors, clustered at the municipality level, are in parentheses.
Significance levels *** p<0.01, ** p<0.05, * p<0.1.

**Table A5.**

Political assassinations of former mayors 2000–21. Characteristics of municipalities during the previous three-year term. Incidence Rate Ratios.

| | (1) Former mayor who left office less than four years ago | (2) Former mayor who left office less than four years ago | (3) Former mayor in politics, business, or an unknown profession, who left office less than four years ago | (4) Former mayor in politics, business, or an unknown profession, who left office less than four years ago | (5) Former mayor who left office less than four years ago | (6) Former mayor who left office less than four years ago | (7) Former mayor in politics, business, or an unknown profession, who left office less than four years ago | (8) Former mayor in politics, business, or an unknown profession, who left office less than four years ago |
|---|---|---|---|---|---|---|---|---|
| | | Fixed Effects Poisson | | | | Second-Stage IV Fixed Effects Poisson | | |
| State actions against illegal drugs (destroyed cultivations, seized drugs, dismantled labs) in the administration where the assassination took place | 1.234 | 1.234 | 2.250* | 2.250* | 7.216 | 6.549 | 2.093 | 1.875 |
| | (0.553) | (0.553) | (0.998) | (0.998) | (25.974) | (23.191) | (10.091) | (8.838) |
| Number of members of organized crime arrested or killed in the administration where the assassination took place | 1.011 | | 1.011 | | 0.817 | | 0.795 | |
| | (0.014) | | (0.013) | | (0.669) | | (0.741) | |
| Number of members of organized crime arrested in the administration where the assassination took place | | 1.012 | | 1.012 | | 0.817 | | 0.796 |
| | | (0.015) | | (0.014) | | (0.668) | | (0.739) |
| Distance to oil pipeline divided by the average price of gasoline and diesel | 0.664*** | 0.664*** | 0.669** | 0.669** | 0.657 | 0.655 | 0.740 | 0.737 |
| | (0.096) | (0.096) | (0.107) | (0.107) | (0.199) | (0.197) | (0.334) | (0.333) |
| Homicide rate in the administration term (excluding criminals and political | 1.001 | 1.001 | 1.000 | 1.000 | 1.002 | 1.002 | 1.002 | 1.002 |
| | (0.001) | (0.001) | (0.002) | (0.002) | (0.006) | (0.006) | (0.007) | (0.007) |
| Political coordination: The municipality was ruled by the same party as its respective state and presidency | 1.184 | 1.187 | 1.470 | 1.476 | 1.240 | 1.200 | 1.649 | 1.589 |
| | (0.622) | (0.622) | (0.882) | (0.884) | (3.007) | (2.889) | (7.527) | (7.248) |
| Annual income of the government municipal budget, in real terms | 1.000** | 1.000** | 1.000* | 1.000* | 1.000 | 1.000 | 1.000 | 1.000 |
| | (0.000) | (0.000) | (0.000) | (0.000) | (0.000) | (0.000) | (0.000) | (0.000) |
| Annual nightlight in logarithm | 0.985 | 0.987 | 1.083 | 1.086 | 0.780 | 0.788 | 0.768 | 0.777 |
| | (0.375) | (0.376) | (0.470) | (0.473) | (0.590) | (0.576) | (0.846) | (0.841) |
| Observations | 275 | 275 | 208 | 208 | 212 | 212 | 150 | 150 |
| Wald Chi2 | 15.91 | 16.10 | 15.43 | 15.71 | 13.59 | 13.59 | 9.006 | 9.006 |
| Log pseudolikelihood | -72.73 | -72.73 | -56.52 | -56.52 | -56.19 | -56.19 | -42.73 | -42.73 |
| Endogeneity test | | | | | 0.560 | 0.540 | 0.340 | 0.340 |
| P-value | | | | | 0.757 | 0.763 | 0.842 | 0.842 |

Robust standard errors, clustered at the municipality level, are in parentheses. IV specifications also have bootstrapped standard errors. First-stage regression is in Table A4. Significance levels *** p<0.01, ** p<0.05, * p<0.1.

**Table A6.**

First-stage regression of Table A7. Panel fixed effects Poisson, Incidence Rate Ratios.

| | (1) Whether the state has taken any actions against illegal drugs | (2) Number of members of organized crime arrested or killed | (3) Whether the state has taken any actions against illegal drugs | (4) Number of members of organized crime arrested or killed |
|---|---|---|---|---|
| | Model 1 | | Model 2 | |
| Distance to the nearest port multiplied by the annual price of cocaine in the United States per kilo | 1.006*** | 1.118*** | 1.006*** | 1.117*** |
| | (0.001) | (0.034) | (0.001) | (0.034) |
| Log Percentage of mountainous territory multiplied by the annual price of heroin in the United States | 0.513*** | 24.982*** | 0.512*** | 24.935*** |
| | (0.044) | (28.661) | (0.044) | (28.482) |
| Mayors, pre-candidates, and candidates assassinated in the administration term prior to the election | 1.058 | 1.822** | | |
| | (0.044) | (0.513) | | |
| Pre-candidates and candidates assassinated in the administration term prior to the election | | | 1.138** | 1.082 |
| | | | (0.068) | (0.251) |
| Distance to oil pipeline divided by the average price of gasoline and diesel | 1.019*** | 1.083 | 1.019*** | 1.075 |
| | (0.006) | (0.055) | (0.006) | (0.055) |
| Homicide rate in the administration (excluding criminals and political assassinations) | 1.000*** | 1.005*** | 1.000*** | 1.006*** |
| | (0.000) | (0.001) | (0.000) | (0.001) |
| Annual income of the government municipal budget, in real terms | 1.000*** | 1.000** | 1.000*** | 1.000** |
| | (0.000) | (0.000) | (0.000) | (0.000) |
| Annual nightlight in logarithm | 1.156*** | 0.625** | 1.157*** | 0.620** |
| | (0.024) | (0.115) | (0.024) | (0.115) |
| Observations | 7,901 | 2,965 | 7,901 | 2,965 |
| Wald Chi2 | 154.5 | 83.04 | 155.3 | 78.13 |
| Log pseudolikelihood | -4842 | -5537 | -4842 | -5566 |
| Excluded instruments | | | | |
| Wald Chi2 | 62.93 | 15.97 | 62.93 | 15.97 |
| Log pseudolikelihood | -5639 | -6309 | -5639 | -6309 |

Robust standard errors, clustered at the municipality level, are in parentheses.

Significance levels *** p<0.01, ** p<0.05, * p<0.1.

**Table A7.**

Turnout and re-election of incumbent party 2000–21. Panel fixed effects.

| | (1) Turnout | (2) Turnout | (3) Incumbent party gets re-elected | (4) Incumbent party gets re-elected | (5) Turnout | (6) Turnout | (7) Incumbent party gets re-elected | (8) Incumbent party gets re-elected |
|---|---|---|---|---|---|---|---|---|
| | Panel Fixed Effects | | | | Second-Stage Panel Fixed Effects IV | | | |
| Whether the state had taken any actions against illegal drugs (destroyed cultivations, seized drugs, dismantled labs) during the administration prior to the election | 0.271 | 0.270 | 0.044*** | 0.044*** | 6.562*** | 6.607*** | 0.402*** | 0.400*** |
| | (0.251) | (0.251) | (0.013) | (0.013) | (1.337) | (1.326) | (0.083) | (0.082) |
| Number of members of organized crime arrested or killed during the administration prior to the election | 0.064*** | 0.063*** | 0.001 | 0.001 | 1.817*** | 1.831*** | -0.010 | -0.010 |
| | (0.018) | (0.018) | (0.001) | (0.001) | (0.160) | (0.161) | (0.009) | (0.009) |
| Number of mayors, pre-candidates, and candidates assassinated in the administration term before the election | -0.740 | | 0.020 | | -1.636 | | 0.008 | |
| | (1.059) | | (0.045) | | (1.045) | | (0.055) | |
| Number of pre-candidates and candidates assassinated in the administration term prior to the election | | -0.618 | | 0.012 | | -1.066 | | -0.029 |
| | | (1.405) | | (0.063) | | (1.494) | | (0.075) |
| Distance to oil pipeline divided by average price of gasoline and diesel | -0.209** | -0.207** | 0.002 | 0.002 | -0.214** | -0.202** | -0.001 | -0.001 |
| | (0.089) | (0.089) | (0.004) | (0.004) | (0.095) | (0.094) | (0.005) | (0.005) |
| Homicide rate in the administration (excluding criminals and political assassinations) | -0.002* | -0.002* | 0.000*** | 0.000*** | -0.018*** | -0.019*** | 0.000 | 0.000 |
| | (0.001) | (0.001) | (0.000) | (0.000) | (0.002) | (0.002) | (0.000) | (0.000) |
| Annual income of the government municipal budget, in real terms | 0.000* | 0.000* | 0.000*** | 0.000*** | -0.000 | -0.000 | -0.000 | -0.000 |
| | (0.000) | (0.000) | (0.000) | (0.000) | (0.000) | (0.000) | (0.000) | (0.000) |
| Annual nightlight in logarithm | -1.096*** | -1.096*** | -0.059*** | -0.059*** | -0.697*** | -0.687*** | -0.085*** | -0.085*** |
| | (0.214) | (0.214) | (0.010) | (0.010) | (0.251) | (0.251) | (0.013) | (0.013) |
| Observations | 10,580 | 10,580 | 10,917 | 10,917 | 8,954 | 8,954 | 9,212 | 9,212 |
| Log pseudolikelihood | -37478 | -37478 | -6005 | -6005 | -31654 | -31654 | -5086 | -5086 |
| Endogeneity test | | | | | 118.540 | 119.780 | 15.150 | 15.140 |
| P-value | | | | | 0.000 | 0.000 | 0.000 | 0.000 |

Robust standard errors, clustered at the municipality level, are in parentheses. IV specifications also have bootstrapped standard errors. First-stage regression is in Table A6 for columns 5-8.

Significance levels *** p<0.01, ** p<0.05, * p<0.1.